\documentclass[11pt]{article} 
\usepackage[paper=a4paper,margin=1.0in]{geometry}

\usepackage{soul}
\usepackage{amssymb,amstext,amsmath,amsthm,mathtools}
\usepackage[dvips]{graphicx}
\usepackage{latexsym}
\usepackage{psfrag}
\usepackage{amsfonts}
\usepackage{bm} 
\usepackage{color}
\usepackage{cite}
\usepackage{hyperref}
\hypersetup{colorlinks,allcolors=blue,citecolor=red,linktocpage=true}
\usepackage[utf8]{inputenc} 
\usepackage{enumitem}
\numberwithin{equation}{section}

\newcommand{\p}{\partial}

\def\a{\alpha}
\def\b{\beta}
\def\g{\gamma}
\def\d{\delta}
\def\D{\Delta}

\def\e{\epsilon}

\def\bra{\langle}
\def\ket{\rangle}
\def\l{\left}
\def\r{\right}
\def\f{\frac}
\def\MO{\mathcal {O}}

\def\ME{\mathcal {E}}

\begin{document}
\title{4D Weyl Anomaly and Diversity of the Interior Structure of Quantum Black Hole}
\author{
{Pei-Ming Ho$^{a,b}$\footnote{pmho@ntu.edu.tw},~
Hikaru Kawai$^{a}$\footnote{hikarukawai@phys.ntu.edu.tw},~
Henry Liao$^a$\footnote{henryliao.physics@gmail.com}~
and Yuki Yokokura$^{c}$\footnote{yuki.yokokura@riken.jp}}
\\
{\small ${}^a$\emph{Department of Physics and Center for Theoretical Physics, 
National Taiwan University, Taipei 106, Taiwan}}\\
{\small ${}^b$\emph{Physics Division, National Center for Theoretical Sciences, Taipei 10617, Taiwan}}\\
{\small ${}^c$\emph{iTHEMS Program, RIKEN, Wako, Saitama 351-0198, Japan}}\\
}

\maketitle
\begin{abstract}

We study the interior metric of 4D spherically symmetric static black holes by using the semi-classical Einstein equation and find a consistent class of geometries with large curvatures. 
We approximate the matter fields by conformal fields and consider the contribution of the 4D Weyl anomaly, giving a state-independent constraint. 
Combining this with an equation of state yields an equation that determines the interior geometry completely. 
We explore the solution space of the equation in a non-perturbative manner for $\hbar$. 
First, we find four types of asymptotic behaviors and examine the general features of the solutions. 
Then, by imposing physical conditions, we obtain approximately a general class of interior geometries: various combinations of dilute and dense structures without a horizon or singularity. 
This represents the diversity of the interior structure. 
Finally, we show that the number of possible patterns of such interior geometries corresponds to the Bekenstein-Hawking entropy.

\end{abstract}

\section{Introduction.}\label{s:intro}

Black-hole geometries have been extensively studied, including quantum effects such as particle creation and vacuum polarization \cite{BD,Parker:2009uva,Hu:2020luk}. 
These quantum effects can develop nontrivial energy-momentum tensors \cite{BD,Parker:2009uva,Hu:2020luk,Robinson:2005pd,Das:2007ru}, and back-react to the background geometry through the semi-classical Einstein equation.
When the energy-momentum tensor is small, we can solve the equation perturbatively \cite{Beltran-Palau:2022nec,Shafiee:2022jfx,Barcelo:2007yk,BD,Parker:2009uva,Hu:2020luk,Arrechea:2021xkp}.
However, it is not possible to have this perturbative scheme in cases where a large energy-momentum tensor is involved. 
In such a non-perturbative regime, some examples of self-consistent solutions can be found in the literature \cite{KMY,KY3,KY4,KY5,Beltran-Palau:2022nec,Arrechea:2021xkp,Y1,KY2,KY1}. 
For example, a large back-reaction can deform event horizons into different geometries: 
wormhole-like structures, geometries with neither a horizon nor a wormhole-like neck, or more \cite{Arrechea:2021ldl,Ho2D1,Ho2D2,Arrechea:2021vqj,Calza:2022szy}. 
In this sense, the geometry inside and near black holes is yet to be determined. 


To explore it in an effective field theory, we consider a 4D spherical static black hole of mass $M\equiv \f{a}{2G} \gg m_P \equiv \sqrt{\hbar/G}$ and construct the interior metric by solving self-consistently the semi-classical Einstein equation
\begin{equation}\label{Einstein}
G_{\mu\nu}=8\pi G \bra \psi |T_{\mu\nu}|\psi\ket,
\end{equation}
where gravity is described by a classical metric $g_{\mu\nu}$, and matter fields by quantum operators. 
At a given point inside an object, if the typical length scale is $L$, we have from~\eqref{Einstein} that $L^{-2}\sim G \hbar L^{-4}$. 
Then, there are two possible length scales: $L=\infty$ and $L\sim\sqrt{G\hbar}\equiv l_P$.
If the length scale is $L=\infty$ everywhere, the object cannot be a black hole, as we usually expect. On the other hand, if $L\sim l_P$ in a certain region, we can indeed construct a black hole-like object~\cite{KY5}.
Therefore, the interior of a black hole is expected to be highly curved in a certain region, and the curvature excites the quantum fields to satisfy \eqref{Einstein}.
The back-reaction from such excited fields, which can be large, determines the interior geometry with high curvatures/densities self-consistently.
On the other hand, the region outside such a highly curved one has smaller curvatures and has little back-reaction from the fields.
As a result of the dynamics governed by \eqref{Einstein}, the boundary between the dense and dilute regions, the surface, should appear at $r=r_{sur}\approx a$, and the black hole should be described as a compact spacetime region with an excited state $|\psi\ket$ consistent with the area entropy and Hawking temperature~\cite{Hawking,Bekenstein}.
In this way, we characterize a black hole not by the existence of horizons but as a compact object whose exterior is approximated by the Schwarzschild metric outside a radius close to the Schwarzschild radius.

Specifically, we consider conformal matter fields in the interior. 
This should be a good approximation because, at least, some regions of the interior have high (near Planckian) curvature and high energy. 
Then, the trace part of the energy-momentum tensor is given by the 4D Weyl anomaly,
\begin{equation}\label{anomaly}
\bra \psi |T^\mu{}_{\mu}|\psi \ket=\hbar \l(c_W  C_{\a\b\g\d}C^{\a\b\g\d}-a_W{\cal G}+b_W \Box R \r),
\end{equation}
independently of the state $|\psi\ket$, 
where $C_{\a\b\g\d}C^{\a\b\g\d}=R_{\a\b\g\d}R^{\a\b\g\d}-2R_{\a\b}R^{\a\b}+\f{1}{3}R^2$ 
and ${\mathcal G}\equiv R_{\a\b\g\d}R^{\a\b\g\d}-4R_{\a\b}R^{\a\b}+R^2$ \cite{BD,Duff,Parker:2009uva,Hu:2020luk}.
Here, the coefficients $c_W$ and $a_W$ are fixed by the matter content of the theory for small coupling constants, 
while $b_W$ depends on the coefficients of the $R^2$ and $R_{\mu\nu}R^{\mu\nu}$ terms of the gravity action.
Using \eqref{anomaly}, the trace of \eqref{Einstein} is a fully geometric equation, 
which is independent of $|\psi \ket$, 
and therefore, any solution to \eqref{Einstein} must satisfy it. 
Note that the Weyl anomaly has been considered in the study of black-hole geometries in previous works~\cite{Christensen:1977jc,Davies:1977bgr,Balbinot:1999ri,Balbinot:1999vg,Mottola:2006ew,Anderson:2007eu,Cai:2009ua,Cognola:2013fva,Cai:2014jea,Abedi:2015yga,Mottola:2016mpl,Ho:2018fwq,Mottola:2022tcn,Beltran-Palau:2022nec,Calza:2022szy,Tsujikawa:2023egy,Fernandes:2023vux}.



Since a 4D spherical static metric has two functional degrees of freedom, we need one more equation to determine the interior. 
We here study three candidates of equations of state (see Appendix \ref{A:whyeos}.), and employ a natural one that is valid when the energy-momentum tensor is large but not necessarily isotropic. 
Then, the metric can be expressed in terms of a single function, $a(r)$, 
where $M(r)\equiv \f{a(r)}{2G}$ is the contribution of the ADM energy of the part within radius $r$. 
Thus, the trace part of \eqref{Einstein} becomes the equation that determines the interior geometry completely, which we call ``\textit{the self-consistent equation for $a(r)$}''. 
It is a nonlinear ordinary differential equation for $a(r)$, involving up to the fourth-order derivative (Sec.~\ref{s:master}). 

The main goal of this paper is to identify the typical geometry -- independent of $b_W$ (or $\beta$, see Sec.~\ref{s:master}) -- of the quantum black hole permitted by the self-consistent equation for $a(r)$.
At the same time, we provide detailed analysis of the self-consistent equation for $a(r)$ in the appendices~\ref{A:full} -- \ref{A:f}.
We would also like to point out that the geometries of black holes we obtain can be non-perturbative in that they do not have an $\hbar \rightarrow 0$ limit.


In Sec.~\ref{s:asympt}, we start by classifying solutions to the self-consistent equation for $a(r)$ and construct the solutions as a (formal) series expansion of $1/r$ at $r\gg l_P$.
Since the equation contains higher derivative terms when $\beta\neq 0$, we need extra care due to the existence of spurious solutions, which are sensitive to $\beta$.
Fortunately, as we demand regular behaviors for solutions, we find that they are not spurious.
In the main text, we only post those relevant to our main result.\footnote{See Appendix~\ref{A:full} for detailed calculations and more general classification.}
They are those close to the Schwarzschild metric and those with high energy density 
and near-Planckian curvature.
Also, we discuss in detail of why these two classes of solutions are valid.

In Appendix~\ref{s:structure}, we numerically solve the self-consistent equation for $a(r)$ and study the relationship between the classified solutions to obtain the global structure of the solution space. From this and Sec.~\ref{s:asympt}, we pick out only the typical interior structure of quantum black holes in Sec.~\ref{s:interior}.
We then employ two physical conditions for the black hole: there is no singularity anywhere, and the interior metric connects to the exterior Schwarzschild metric at $r=r_{sur}\approx a$.
Under these conditions, for any value of $b_W$, a typical physical geometry is obtained.
It starts from a flat region around the center, approaches the high-density metric as $r$ increases, and then connects to the Schwarzschild metric at $r=r_{sur} = a + \MO(\frac{l_P{}^2}{a}) > a$.
Thus, there is no horizon but a surface at $r=r_{sur}$.
This is the simplest interior structure that satisfies the self-consistent equation for $a(r)$ and the physical conditions.
Roughly speaking, this represents a spherical shell with a finite width.

In Sec~\ref{s:entropy}, we discuss more general internal structures of quantum black holes consisting of multiple shells.  That allows
more complex internal structures that satisfy the self-consistent equation for $a(r)$ and physical conditions.
This complexity creates the diversity of the interior geometries, leading to a natural conjecture: this variety, i.e., the possible patterns of the classical geometry of the black hole interior, could explain the origin of black hole entropy.
This is similar to the Fuzzball\cite{Mathur} but differs significantly in that the geometries are interior solutions of the semi-classical Einstein equations \eqref{Einstein}.
Hence, we provide a speculative discussion on this conjecture.
Interestingly, \eqref{Einstein} leads to a minimum length scale where the interior structure can change, which is near the Planck length in the proper radius length.
Using this length and considering the contribution from the matter-field configuration according to Bekenstein’s idea\cite{Bekenstein}, we can show that the entropy corresponding to the number of possible patterns of the classical interior geometry satisfying \eqref{Einstein} agrees with the area law up to a numerical constant.

\section{Setup and the self-consistent equation for \texorpdfstring{$a(r)$}{}}\label{s:master}
In this section, we are going to derive the self-consistent equation for $a(r)$ from~\eqref{Einstein} and an equation of state.
We assume that our metric is static and spherically symmetric and split the region into the interior and exterior parts by $r_{sur}$, a quantity that will be determined by our solution.
We express the interior region ($r \leq r_{sur} $) as 
\begin{equation}\label{metric}
ds^2=-\l(1-\f{a(r)}{r}\r)e^{A(r)}dt^2+\f{1}{1-\f{a(r)}{r}}dr^2+r^2d\Omega^2,
\end{equation}
where as a result of $G_{tt}=8\pi G \bra T_{tt}\ket$, we have \cite{Poisson,Landau_C}
\begin{equation}\label{Mr}
    \f{a(r)}{2G}\equiv M(r)= 4\pi \int^r_0 dr' r'^2 \bra -T^t{}_t(r')\ket.
\end{equation}
The exterior region ($r \geq r_{sur} $) can be approximated as 
\begin{equation}\label{Sch}
    ds^2=-\l(1-\f{a}{r}\r)dt^2+\f{1}{1-\f{a}{r}}dr^2+r^2d\Omega^2, 
\end{equation}
if we ignore a small back-reaction of vacuum polarization for simplicity.\footnote{Note that a large back reaction from quantum effects is considered in the interior metric \eqref{metric}, and the exterior has at most a small one.} 
Here, $\f{a}{2G}\equiv M\equiv \lim_{r\to \infty} M(r)$ is the ADM mass of the system.
These two metrics must be connected continuously at the boundary at $r=r_{sur}\approx a$ (see \eqref{surface} for the precise value), 
which will play a role of a physical condition to determine the interior later (Sec.~\ref{s:interior}). 

Next, we introduce an equation of state to eliminate one functional degree of freedom, leaving only $a(r)$ to be solved. 
We note here Buchdahl's theorem:
There exists no static spherical configuration of isotropic fluid with size $r_{sur}\approx a$ 
and energy density $\bra -T^t{}_t(r)\ket$ satisfying $\bra -T^t{}_t(r)\ket\geq 0$ and $\f{d}{dr}\bra -T^t{}_t(r)\ket\leq0$ \cite{Buchdahl,Wald_text}. 
In our case, we only assume positive energy and pressures, since the physical excitation of the fields in a static gravitational field should be locally consistent with thermodynamics and have positive energy and pressures \cite{Landau_SM}. 
We examine three possibilities for equations of state by analyzing the thermodynamical consistency and regularity of asymptotic behaviors (see Appendix \ref{A:whyeos}.).
We find that 
\begin{equation}\label{eos}
    \bra T^r{}_r\ket=\f{2-\eta}{\eta} \bra -T^t{}_t\ket
\end{equation}
is the only candidate equation of state in dense regions.
Note that it can describe an anisotropic structure, not necessarily an isotropic fluid.
In fact, as we will see in Sec.~\ref{s:interior}, our solutions have large anisotropy, avoiding Buchdahl's theorem.
In particular, we have shell-like solutions as the typical structure, which are  manifestly anisotropic.
Here, because conformal matter fields have no length scale, 
$\eta$ must be constant from dimensional analysis. 
Positivity of $\bra T^r{}_r\ket$ and $\bra -T^t{}_t\ket$ means
\begin{equation}\label{eta}
    0<\eta<2.
\end{equation}
Note that this equation of state~\eqref{eos} with constant $\eta$ is valid only for interior configurations 
where the energy density is so high that the conformal-matter approximation is allowed.
In the following, we assume \eqref{eos} for a given $\eta$ satisfying \eqref{eta}, 
and will check later the self-consistency after obtaining solutions of \eqref{eom} (see Sec.~\ref{s:valid_eos}.).

Now, we are ready to derive the self-consistent equation for $a(r)$. 
We first apply the Einstein equation \eqref{Einstein} to the equation of state \eqref{eos} 
to obtain $G^r{}_r=\f{2-\eta}{\eta} (-G^t{}_t)$. 
For the metric \eqref{metric}, this provides 
\begin{equation}\label{A}
    A(r)=\int^r_{r_0}dr'\f{2\p_{r'} a(r')}{\eta(r'-a(r'))},
\end{equation}
where $r_0$ is a reference point.
Second, we use the anomaly formula \eqref{anomaly} and write the trace part of \eqref{Einstein} as 
\begin{equation}\label{trace}
    G^\mu{}_\mu=\g C_{\a\b\g\d}C^{\a\b\g\d}-\a {\cal G}+ \b \Box R,
\end{equation}
where $\g\equiv 8\pi l_P{}^2 c_W,~\a\equiv 8\pi l_P{}^2 a_W,~\b\equiv 8\pi l_P{}^2 b_W$. 
Finally, we apply the metric \eqref{metric} with \eqref{A} to \eqref{trace} and obtain 
\begin{align}\label{eom}
    \beta a''''(r)
    +\beta \frac{\left(2-\eta \right) \left(3-\eta \right) r a'(r)+\left(\eta +1\right) \eta a(r)-2 \left(\eta -1\right) \eta  r}{\left(2-\eta \right) \eta \left(r-a(r)\right) r}a'''(r)&\nonumber\\
    +\frac{6 \beta -\gamma(2-\eta)}{3 \eta(r-a(r))} a''(r)^2+\mathcal{C}_1(r)a''(r)+\mathcal{C}_2(r)&=0,
\end{align}
where $a''(r)=\f{d^2}{dr^2}a(r)$ and so on, and $\mathcal{C}_1(r)$ and $\mathcal{C}_2(r)$ denote terms containing $a'(r),a(r),r$
(see Appendix \ref{A:eom} for the full expression.). 
This is \textit{the self-consistent equation for $a(r)$}, which we denote as $\ME=0$, 
determining $a(r)$ for a given set $(\eta; \g, \a, \b)$.
Note that the flat space, $a(r)=0$, satisfies \eqref{eom}. 

For a special case where $\b=0$, 
\eqref{eom} becomes quadratic in $a''(r)$. 
Solving it for $a''(r)$ yields
\begin{equation}\label{eom0}
    a''(r)-\frac{1}{2 \gamma  (2-\eta) \eta  r^2 (r-a(r))}
    \left( \mathcal{F}_1(r) \pm  \eta ^{3/2} \left(r-a(r) \right)\sqrt{\mathcal{F}_2(r)} \right)=0
\end{equation}
with 
\begin{align}
\mathcal{F}_1(r)&=
    -2 \gamma  \left(2-\eta \right) r^2 a'(r)^2
    -2 \gamma  \eta  r \left(\left(5-4 \eta \right) a(r)+4 \left(\eta -1\right) r\right) a'(r)\nonumber\\
    &~~~~~~~~~~~~~~~~~~~~~~~~~~~~~~~~~~~~~~~~~~~~~~~~~~~~
    +3 \eta ^2 \left(r-a(r)\right)\left(4 \left(\gamma -\alpha \right)a(r)+r^3\right)\\
\mathcal{F}_2(r)&=-72 \gamma  \left(\eta -1\right) r^4 a'(r)
    -48 \alpha  \gamma  \left(2-\eta\right) r^2 a'(r)^2
    +9 \eta \left(8 \left(\alpha -\gamma \right) a(r) \left(2 \alpha a(r)-r^3\right)+r^6\right). 
\end{align}
This indicates that there are two branches in the solution space of \eqref{eom}. 
We denote the $``+"$ and $``-"$ ones in \eqref{eom0} as $\ME_+=0$ and $\ME_-=0$, respectively. 
Here, let's look at the difference a little. 
By substituting $a(r)=0$, we get $\ME_-=0$ and $\ME_+=-\f{3\eta}{\g(2-\eta)}r$ for $r\geq 0$. 
From this, we can expect that the $\ME_-$ branch represents spacetimes with small curvatures, while the $\ME_+$ branch represents spacetimes with high curvatures.
This will become clear in Sec.~\ref{s:sol} when we discuss the physical properties of solutions.

We finally discuss the coefficients $c_W,a_W,b_W$.
Since here we only discuss unitary matter fields, $a_W$ and $c_W$ are assumed to be positive.
Both are proportional to the total number of degrees of freedom in the theory, $N$.\footnote{The explicit forms of $c_W$ and $b_W$ for free fields are given by 
\begin{equation}
c_W=\f{1}{1920\pi^2}(N_S+6N_F+12N_V),~~~a_W=\f{1}{5760\pi^2}(N_S+11N_F+62N_V),\nonumber
\end{equation}
where $N_S$, $N_F$, and $N_V$ are respectively the numbers of scalars, spin-$\f{1}{2}$ Dirac fermions, 
and vectors in the theory \cite{Duff,BD}. Indeed, this satisfies \eqref{a/c}.} 
Here, the ratio $\f{\a}{\g}=\f{a_W}{c_W}$ satisfies an interesting inequality: 
\begin{equation}\label{a/c}
\f{1}{3}\leq \f{\a}{\g} \leq \f{31}{18}, 
\end{equation}
which holds true even when the fields are interacting \cite{Hofman}. 
On the other hand, $b_W$ depends on the effects of the finite renormalization in the gravity action 
and can be any value, positive, negative, or zero, if the theory is freely chosen. 
In the following, however, we assume that $b_W$ is a constant of the order of $\MO(N)$ that can be zero depending on the theory.\footnote{
This should be natural because when renormalizing $\bra\psi|T_{\mu\nu}|\psi\ket$ in a theory with $N$ degrees of freedom, a finite renormalization proportional to $N$ is often involved since the counter terms for the UV divergence are proportional to $N$ \cite{BD}. See \cite{KY4} for an example. }
Thus, we have 
\begin{equation}\label{abc}
\g,\a,\b\sim N l_P{}^2.
\end{equation}

\section{Asymptotic solutions}\label{s:asympt}




In this section, we classify the solutions of the equation of motion~\eqref{eom} in terms of series expansion. Here, only those solutions that are not spurious, i.e., those with a weak dependence on $\beta$, will be presented that are relevant to the main text.

First, in Sec.~\ref{s:sol}, we obtain the non-spurious asymptotic solution of~\eqref{eom} as a series expansion in the limit of $r\gg l_P$. This requires the following steps: (1) find all possible leading powers of the series (2) remove all solutions with irregular behavior, (3) complete the remaining series as asymptotic solutions. We then list the solutions needed for later discussion, along with their physical properties. See Appendix~\ref{A:full} for a complete classification of solutions.

Sec.~\ref{s:valid_eos} examines the validity of these solutions, along with the validity of the equation of state~\eqref{eos}.

\subsection{Relevant solution series}\label{s:sol}

As a first step, we set $a(r)\sim r^k$, expand \eqref{eom} for $r\gg l_P$, and pick up all possible leading terms (for a given $k$).
We can find that their powers are $k+5, 2k+4, 3k+3, 4k$. 
Then, we require the leading terms to be zero, which can be satisfied in two ways: a single leading term with a vanishing coefficient, and multiple leading terms with their coefficients canceling each other.

In the case of a single leading term, one can show that it can only happen when $k\geq 3$ (see Appendix \ref{A:k}), and $a(r)$ either hits $a=r$ singularity\footnote{$a=r$ singularity means that curvature invariants are singular when $a(r)=r$. For example, scalar curvature is inversely proportional to $(r-a(r))^2$.} quickly or goes to $-\infty$ as $r$ grows.
Hence, we discard these possibilities to avoid irregular behaviors.

\begin{figure}[!ht] 
    \centering
    \includegraphics[scale =0.7]{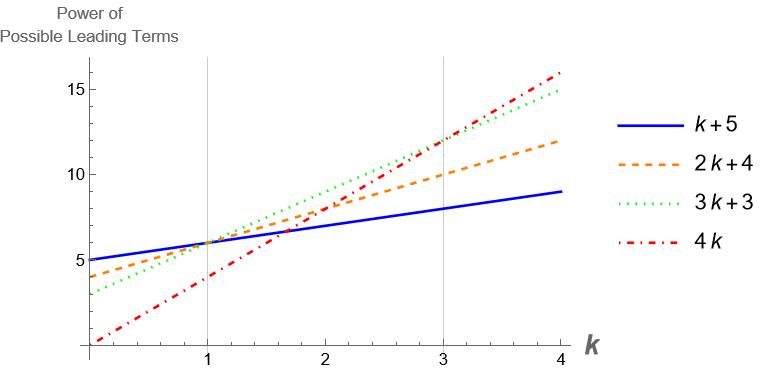}
    \caption{The powers of the leading terms in \eqref{eom} for $a(r)\sim r^k$.}
    \label{f:k_plot}
\end{figure}

As for the other possibility, we can focus on the intersections among the leading powers in terms of $k$.
Plotting $k+5, 2k+4, 3k+3, 4k$ as in Fig.~\ref{f:k_plot}, 
we can see easily that at $k=1$ three lines cross with the maximum power 6, and at $k=3$ two lines cross with the maximum power 12. 
For $k\neq0$, therefore, the leading behaviors are $a(r)\sim r$ and $a(r)\sim r^3$. However we do not keep $a(r)\sim r^3$ in the main text, since it has irregular behavior and also requires large negative energy (see discussion around~\eqref{neg_energy} for more details).

For $k=0$, we need extra care because it vanishes when differentiated. 
Then, setting $a(r) = C + r^n$ and using the same procedure as above to find the sub-leading terms, we can see that only $n=-3$ is possible. 

Thus, we have obtained the relevant possible leading solutions of \eqref{eom}: $a(r)=\MO(C+r^{-3}),~\MO(r)$.
In the following, we mention only the relevant ones to our main results: the dilute density solution $a_S(r)$ and the high density solution $a_{den}(r)$.

Note that we perform detailed analysis on constructing 4 types of solution series in Appendix~\ref{A:full}.
Among them, we find that $\beta$-corrections never come in as leading order in $a(r)$. That means, in the limit of $\beta \rightarrow 0$, these $a(r)$ do not become singular, and therefore, are not artifacts of higher order derivatives in~\eqref{eom}, or equivalently, are not spurious.

\subsubsection*{Dilute density solution $a_S(r)$}

From the leading behavior $a(r)=\MO(C+r^{-3})$, we can construct the $a_S(r)$ solution. 
The series of $a_S(r)$ with an unfixed constant $c_0$ is given as (see Appendix~\ref{A:full} for the derivation)
\begin{equation}\label{a_S}
a_S(r)=c_0 +\frac{2 {c_0}^2 \eta  (\alpha -\gamma )}{(\eta -3) r^3}
+\frac{3 {c_0}^3 \eta  (\alpha -\gamma )}{2(\eta -3)(8-3 \eta ) r^4}
+\frac{9 {c_0}^2 \eta  (\alpha -\gamma ) \left(-20 \beta  (8-3 \eta)+{c_0}^2\right)}{5 (2 \eta -5) (8-3 \eta) r^5}+\MO(r^{-6}),
\end{equation}
which is a deformation of the Schwarzschild metric of mass $\f{c_0}{2G}$.

From \eqref{metric} with \eqref{A}, the leading behavior of the metric for \eqref{a_S} becomes 
\begin{equation}\label{dilute}
ds^2\approx-\l(1-\f{c_0}{r}+\frac{2 \eta  (\alpha-\g ){c_0}^2}{(3-\eta) r^4}\r)
e^{-\frac{3\eta  (\alpha-\g ){c_0}^2}{(3-\eta) r^4}}dt^2
+\l(1-\f{c_0}{r}+\frac{2\eta  (\alpha-\g ){c_0}^2}{(3-\eta) r^4}\r)^{-1}dr^2+r^2d\Omega^2,
\end{equation}
where we have absorbed the integration constant from \eqref{A} to the redefinition of $t$ and kept the leading correction.
This metric is controlled by a length-dimensional constant $c_0$. 
Also, it is perturbative in $\hbar$ since it is still finite as $\hbar\rightarrow 0$.

We can calculate the Einstein tensors as
\begin{equation}
-G^t{}_t=\frac{6 \eta  (\alpha-\g ){c_0}^2}{(3-\eta) r^6},~
G^r{}_r=\frac{6 (2-\eta)(\alpha-\g ){c_0}^2}{(3-\eta) r^6},~
G^\theta{}_\theta=-\frac{12 (2-\eta) (\alpha-\g ){c_0}^2}{(3-\eta) r^6},
\end{equation}
with the sub-leading terms of $\MO(r^{-7})$, and the curvatures as 
\begin{align}
R=\frac{12 (\alpha -\gamma )c_0^2}{r^6}+\MO(r^{-8}),&~~
R_{\mu\nu\a\b}R^{\mu\nu\a\b}
=\frac{12 {c_0}^2}{r^6}+\frac{48 (6-5 \eta) (\alpha -\gamma )c_0^3}{(3-\eta) r^9}+\MO(r^{-10}),\nonumber\\
R_{\mu\nu}R^{\mu\nu}&=\frac{72 \left(5 \eta ^2-18 \eta +18\right) (\alpha -\gamma )^2c_0^4}{(3-\eta)^2 r^{12}}+\MO(r^{-13}).
\end{align}
For \eqref{eta} and \eqref{a/c}, the energy density $\bra -T^t{}_t\ket=\f{1}{8\pi G} (-G^t{}_t)$ can be positive or negative depending on the ratio $\a/\g$. 
For $c_0\sim r \gg l_P$ (at most), the energy density and curvatures are small compared to $\MO(1)$ (from \eqref{abc}). 
Therefore, \eqref{dilute} represents a modified Schwarzschild metric with small quantum corrections.

\subsubsection*{High density solution $a_{den}(r)$}

For $a(r)=\MO(r)$, we construct the $a_{den}(r)$ solution, which takes the form
\begin{align}\label{a_den}
&a_{den}(r)=r-\frac{2 \gamma }{3 \eta ^2 r}+\frac{4 \gamma  \left(\gamma  \left(\eta ^2+6 \eta -2\right)-6 \alpha  \eta ^2\right)}{9 \eta ^4 r^3}\nonumber\\
&-\frac{8 \gamma  \left(-6 \gamma  \eta ^2 \left(\alpha  \left(\eta ^2+15 \eta -7\right)-3 \beta \right)+18 \alpha  \eta ^3 (2 \alpha  \eta -\beta )+\gamma ^2 \left(\eta ^4+13 \eta ^3+55 \eta ^2-58 \eta +13\right)\right)}{27 \eta ^6 r^5} \nonumber \\
&+\MO(r^{-7}),
\end{align}
which represents a dense interior in the sense that $a_{den}(r)\approx r$. 

The leading behavior of the metric for \eqref{a_den} is 
\begin{equation}\label{dense}
ds^2\approx-\frac{2 \gamma }{3 \eta ^2 r^2}e^{\frac{3 \eta  r^2}{2 \gamma }}dt^2+\frac{3 \eta ^2 r^2}{2 \gamma }dr^2+r^2 d\Omega^2,
\end{equation}
which depends only on $\g$ and $\eta$. This has 
\begin{equation}\label{G_den}
-G^t{}_t=\f{1}{r^2}+\MO(r^{-4}),~
G^r{}_r=\frac{2-\eta}{\eta}\f{1}{r^2}+\MO(r^{-4}),~
G^\theta{}_\theta=\frac{3}{2\g}+\MO(r^{-2})
\end{equation}
and 
\begin{equation}\label{R_den}
R=-\f{3}{\g},~~R_{\mu\nu\a\b}R^{\mu\nu\a\b}=\f{9}{\g^2},~~R_{\mu\nu}R^{\mu\nu}=\f{9}{2\g^2},
\end{equation}
with the sub-leading corrections of $\MO(r^{-2})$. 

First, the metric is non-perturbative for $\hbar$ since we cannot take the limit $\hbar\to0$. 
Second, it is uniform and almost Planckian in that the curvature invariants are constants of $\MO\l(\f{1}{N l_P{}^2}\r)$, which is still smaller than the Planck scale for a large $N$. 
In this sense, this represents a semi-classical, dense configuration.
Third, from \eqref{R_den}, we have $R=-\f{2}{L^2}+\MO(r^{-2})$, which indicates that the metric is a warped product of $AdS_2$ of radius $L\equiv\sqrt{\f{2\g}{3}}\sim \sqrt{N}l_P$ and $S^2$ of radius $r$.\footnote{\label{foot:AdS}More explicitly, we can check it as follows \cite{KY4,KY5}. 
In general, $ds^2=-f(r)dt^2+h(r)dr^2+r^2d\Omega^2=\f{L^2}{z^2}(-dt^2+dz^2)+r(z)^2d\Omega^2$ holds if the condition $\sqrt{h(r)}=\pm\f{L}{2}\p_r \log f(r)$ is satisfied. 
Using \eqref{dense} and focusing a point $r$, we have $h(r)=\frac{3 \eta ^2 r^2}{2 \gamma }$ and $\p_r \log f(r)\approx \p_r A(r)=\frac{3 \eta  r}{\gamma }$, and then the condition holds for $L=\sqrt{\f{2\g}{3}}$.} 
Fourth, the tangential pressure $\bra T^\theta{}_\theta\ket$ is so large that $-G^t{}_t,G^r{}_r \ll G^\theta{}_\theta$, breaking the dominant energy condition and making the interior anisotropic drastically.
This comes from the vacuum fluctuations of all the modes with any angular momentum \cite{KY4}. 
Fifth, the energy density and pressures are positive for \eqref{eta}, 
and thus this metric can represent a physical excitation.

\subsection{Validity of the equation of state}\label{s:valid_eos}
We here discuss the validity of the equation of state \eqref{eos} for the asymptotic solutions above.

We first study $a_{den}(r)$. 
The curvatures \eqref{R_den} are large, 
and the characteristic energy scale is close to the Planck scale. 
For such a high-energy region, 
the equation of state \eqref{eos} (with a constant parameter $\eta$ satisfying \eqref{eta}) should work well. 
In fact, there are several pieces of evidences supporting this expectation. 
First, the dense metric \eqref{dense} can be obtained 
by considering a spherical collapsing matter as a collection of many concentric shells 
and analyzing the time evolution self-consistently with \eqref{Einstein} \cite{KMY,KY2,KY1,KY4,Ho1,Ho2,Ho3}.
In particular, we can show that Hawking radiation occurs in the time-dependent geometry of the dense configuration (see \cite{KMY,KY4}.).
In \cite{Ho3}, $\eta$ can be understood as a parameter characterizing the velocity distribution of the shells. 
In \cite{KY1}, $\eta$ plays a role of an effective parameter representing interactions/scattering 
of the particles inside black holes. 
Second, we can determine the value of $\eta$ by solving \eqref{Einstein} self-consistently\cite{KY4} as follows. 
We first solve the equation of motion for matter fields on the background metric \eqref{dense}; 
we then use the solutions and evaluate the renormalized energy-momentum tensor $\bra\psi|T_{\mu\nu}|\psi\ket$;
we finally equate both sides of \eqref{Einstein} to obtain the self-consistent value of $\eta$. 
Thus, we can justify the use of the equation of state \eqref{eos} for $a_{den}(r)$. 

Next, we consider solutions with low energy density, such as $a_{S}(r)$.
They don't have large curvatures, 
and the equation of state \eqref{eos} may not be valid for them. 
Nevertheless, $a_S(r)$ is an interesting example, since it gives an example of modified Schwarzschild solutions with small quantum correction.
In particular, 
such modified Schwarzschild solutions with small quantum corrections as $a_S(r)$ 
can be obtained for various types of equations of state (see Appendix \ref{a:eos_sch}.) 
and should exist universally for most equations of state. 
Therefore, in the following sections, we assume that 
$a_S(r)$ and the Schwarzschild metric without modification 
are good approximate solutions of the semi-classical Einstein equation \eqref{Einstein}
with the true (but still unknown) equation of state.


\section{Simplest typical geometries of quantum black holes}
\label{s:interior}


In Sec.~\ref{s:sol}, we found asymptotic solutions $a_S(r)$~\eqref{a_S} and $a_{den}(r)$~\eqref{a_den} to the equation ${\ME}=0$~\eqref{eom} as $1/r$ expansions. 
In general, a given solution to ${\ME}=0$ may be described by one asymptotic solution at small $r$ and ``evolve'' to another asymptotic solution at large $r$. 
All possible interpolations of asymptotic solutions admissible by ${\ME}=0$ are studied in Appendix~\ref{s:structure}. 
It is found that generic solutions relevant to black holes have the common behavior that they are well approximated by $a_S(r)$ at small $r$ but by $a_{den}(r)$ at large $r$, for any value of $b_W$. 
This is the simplest one in the context of the diversity of the interior geometries (see Sec.~\ref{s:entropy}). 

In Sec.~\ref{s:typical_interior}, we explain briefly how this typical solution that interpolates between $a_S(r)$ and $a_{den}(r)$ can be constructed. 
In Sec.~\ref{s:physical_structure}, we impose physical conditions on them to obtain physically sound solutions for quantum black holes.


\subsection{Simplest typical interior structures}
\label{s:typical_interior}

We show how the simplest solutions that are described by $a_S(r)$ at small $r$ and interpolate to $a_{den}(r)$ at large $r$ appears regardless of whether $\beta$ vanishes.

\begin{figure}[!ht]
    \centering
    \includegraphics[width=0.4\textwidth]{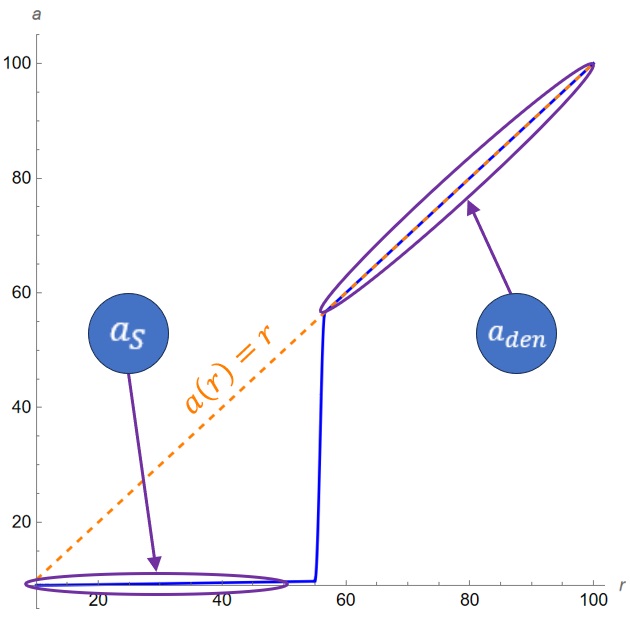}
    \caption{Simplest typical behavior of $a(r)$ for $\beta=0$ (Artificial).}
    \label{f:b0}
\end{figure}

In the case of $\beta=0$, \eqref{eom} reduces to \eqref{eom0}.
One can check with direct substitution of $a_S(r)$ and $a_{den}(r)$ into \eqref{eom0} to see that $a_S(r)$ belongs to $\ME_-=0$, while $a_{den}(r)$ belongs to $\ME_+=0$.
To construct the simplest structure, we can solve \eqref{eom0} in a domain of $r$, say from $r_1$ to $r_3$, and arbitrarily select an $r_2$ in between.
Then, given an initial condition suitable for $a_S(r)$ at $r_1$, we solve $\ME_-=0$ from $r_1$ to $r_2$, and then switch to solving $\ME_+=0$ from $r_2$ to $r_3$ such that $a(r)$ is differentiable at $r_2$.
The resulting solution exhibits the typical geometric structure mentioned above.
See Fig.~\ref{f:b0}.

\begin{figure}[!ht]
  \begin{minipage}[!ht]{0.45\linewidth}
    \centering
    \includegraphics[keepaspectratio, scale=0.5]{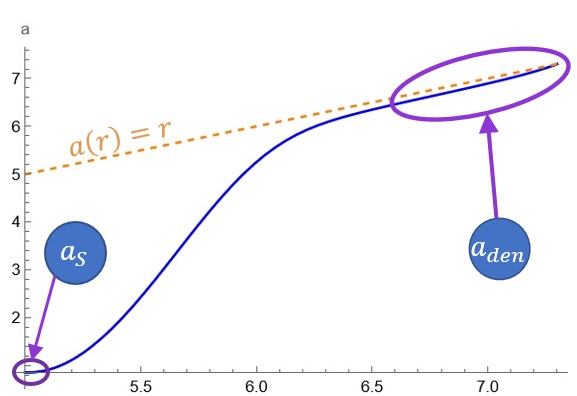}
    \caption{Simplest typical behavior of $a(r)$ for $\beta>0$ (Flow (d)).}
    \label{f:d4}
  \end{minipage}
  \hfill
  \begin{minipage}[!ht]{0.45\linewidth}
    \centering
    \includegraphics[keepaspectratio, scale=0.5]{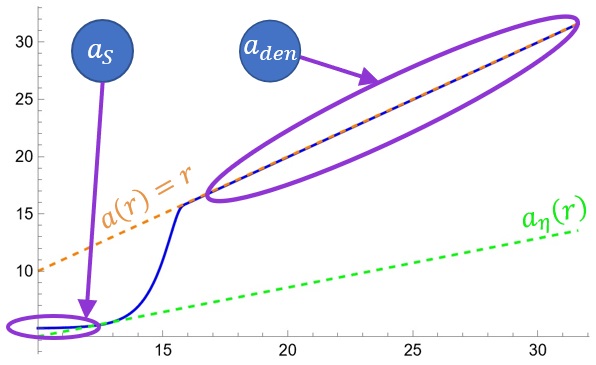}
    \caption{Simplest typical behavior of $a(r)$ for $\beta<0$ (Flow (h)).}
    \label{f:h4}
  \end{minipage}
\end{figure}

\begin{figure}[!ht]
    \centering
    \includegraphics[keepaspectratio, scale=0.7]{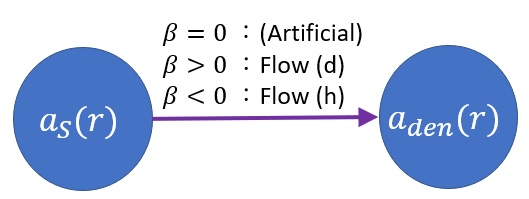}
    \caption{Simplest typical interior structure in the solution space of \eqref{eom}. The phrase ``flow (d)'' and ``flow (h)'' refer to discussions in Appendix~\ref{s:structure}. See Figs.~\ref{f:d} and~\ref{f:h}.}
    \label{f:main_flow}
\end{figure}

As for $\beta\neq 0$, one can numerically verify that the solution with the simplest behavior indeed exists (see Figs.~\ref{f:d4} and~\ref{f:h4}.).
Here, note that due to the higher-order derivative terms associated with $\beta\neq 0$, solutions belonging to different branches are automatically connected.
One can check Appendix~\ref{s:structure} for details of creating these examples, and why there is no other typical interior structure for a quantum black hole according to~\eqref{eom}.
We summarize the typical solutions in Fig.~\ref{f:main_flow}.

\subsection{Physical constraints on black-hole geometry}
\label{s:physical_structure}
Now, we extend the simplest typical interior structure mentioned above to study the entire geometry of a black hole with mass $\frac{a_0}{2G}$.
We impose two physical conditions. 
First, there should be no singularity anywhere (condition 1). 
Second, 
the interior metric is continuously connected to the exterior Schwarzschild metric \eqref{Sch} with the mass $\f{a_0}{2G}$ 
at the surface $r=r_{sur}\approx a_0$ (condition 2).

These two conditions play the role of boundary conditions for the solutions. 
First, condition 1 requires that the region near $r=0$ is well-approximated by a flat spacetime, 
since a point source at $r = 0$ would be singular. 
This can be easily realized by $a_S(r)$ \eqref{a_S} with $c_0 = 0$. 
Condition 2 indicates that the region just below the surface should be well approximated by $a_{den}(r)$, because only $a_{den}(r)$, \eqref{a_den}, can be connected to the Schwarzschild metric near the Schwarzschild radius. (as we will see below).
As a simple prescription, 
$r_{sur}$ can be determined by equating $a_0$ to $a_{den}(r)$ at $r=r_{sur}$:
\begin{align}\label{surface}
&a_0=a_{den}(r_{sur})\approx r_{sur}-\f{2\g}{3\eta^2r_{sur}}\nonumber\\
&\Longrightarrow \ \ r_{sur}\approx a_0+\f{2\g}{3\eta^2 a_0},
\end{align}
where the second term is of the order of $\MO\l(\f{Nl_P{}^2}{a_0}\r)$ from \eqref{eta} and \eqref{abc}. 

The reason why we need $a(r)$ to be approximated by $a_{den}(r)$ under the surface at $r_{sur}$ is as follows. 
It might appear at first sight that $a_S(r)$ with $c_0=a_0$ can satisfy condition 2, 
but it violates condition 1, since it becomes singular at small $r$, as in Fig.~\ref{f:g}.
Thus, a solution starting from $a_{S}(r)$ with $c_0 \approx 0$ around $l_P$\footnote{Since we are solving a semi-classical theory, we cannot have reliable prediction around $l_P$.} and approaching to $a_{den}(r)$ around $r=r_{sur}$ with the Schwarzschild metric for $r>r_{sur}$ describes a physically sound black hole solution.
It has the simplest typical interior geometry mentioned in Sec.~\ref{s:typical_interior} with $c_0\approx 0$ at small $r$.
Note that the total size~\eqref{surface} surpasses Buchdahl’s bound, which originates from the large tangential pressure~\eqref{G_den} in $a_{den}(r)$.

The remaining tasks to build a physical structure are to check (1) whether $c_0\approx 0$ at small $r$ is actually allowed and (2) what is the suitable junction condition for us to connect the interior structure to the exterior Schwarzschild metric.

\begin{figure}[!ht]
  \begin{minipage}[!ht]{0.3\textwidth}
    \centering
    \includegraphics[keepaspectratio, scale=0.3]{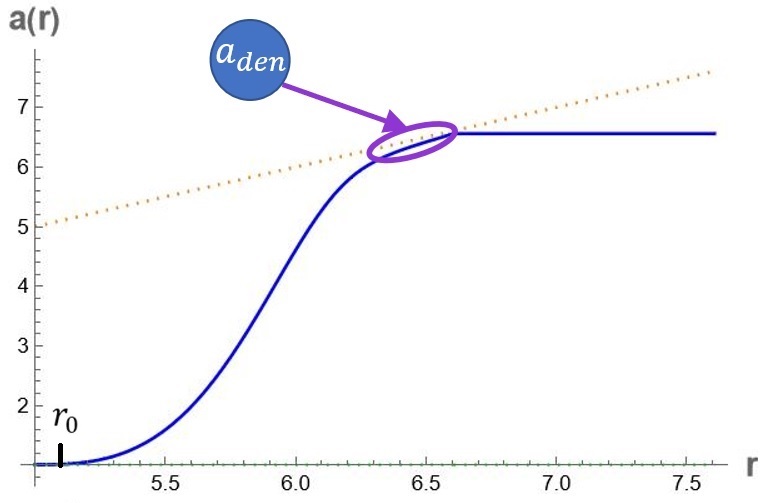}
    \caption{Simplest typical geometry of black hole with $\beta >0$.}
    \label{f:d_to_sch}
  \end{minipage}
  \hfill
  \begin{minipage}[!ht]{0.3\textwidth}
    \centering
    \includegraphics[keepaspectratio, scale=0.3]{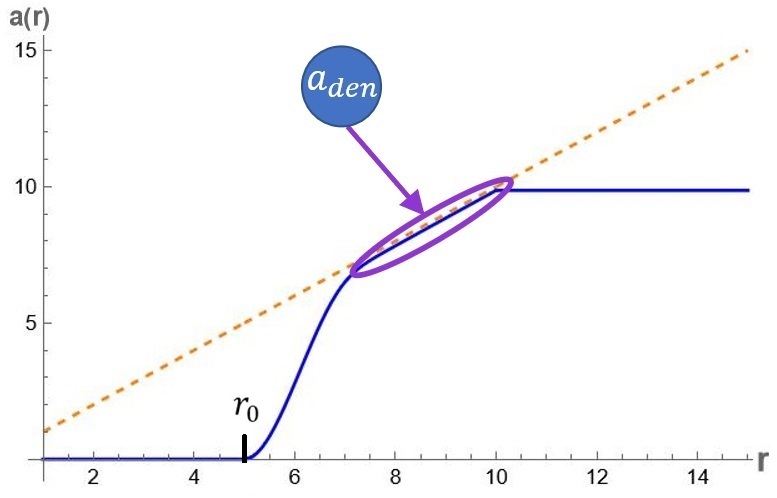}
    \caption{Simplest typical geometry of black hole with $\beta =0$.}
    \label{f:br1_br2_sch}
  \end{minipage}
  \hfill
  \begin{minipage}[!ht]{0.3\textwidth}
    \centering
    \includegraphics[keepaspectratio, scale=0.35]{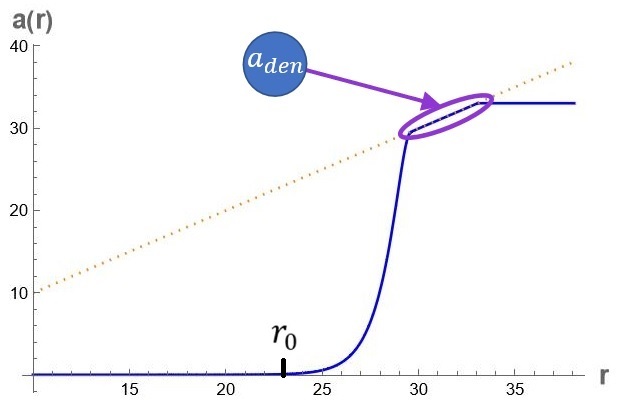}
    \caption{Simplest typical geometry of black hole with $\beta <0$.}
    \label{f:h_to_sch}
  \end{minipage}
\end{figure}

\begin{figure}[!ht]
    \begin{minipage}[!ht]{0.45\linewidth}
        \centering
        \includegraphics[keepaspectratio, scale=0.5]{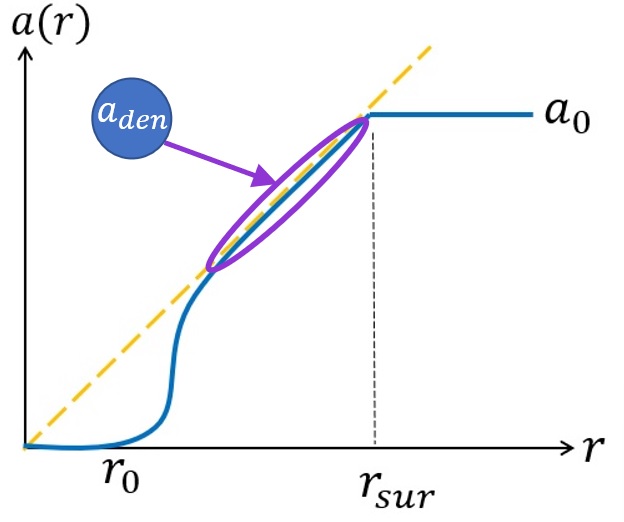}
        \caption{Plot of $a(r)$ obtained for any value of $\beta$.}
        \label{f:uni_a}
    \end{minipage}
    \hfill
    \begin{minipage}[!ht]{0.45\linewidth}
        \centering
        \includegraphics[keepaspectratio, scale=0.7]{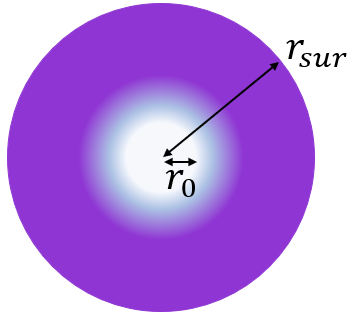}
        \caption{The typical interior configuration represented by Fig.~\ref{f:uni_a}.}
        \label{f:uni_pic}
    \end{minipage}
\end{figure}

For the first task, we give numerical examples for different values of $\beta$ in Figs.~\ref{f:d_to_sch} --~\ref{f:h_to_sch}.
To create these examples, we use initial conditions close to $a_S$ at small $r$ and integrate through Eqs.~\eqref{eom} or~\eqref{eom0} depending on the value of $\beta$. 
The labels $r_0$ in the figures are used to denote the position where the deviation from $a_S$ becomes large as $r$ increases.\footnote{In principle, $r_0$ could be determined by the time evolution of the system, and it will be discussed in future work.}
Hence the lines shown in Figs.~\ref{f:d_to_sch} --~\ref{f:h_to_sch} are integrated results.
Note that, in the case of $\beta=0$, we have $a_S$ in $\ME_-=0$ but $a_{den}$ in $\ME_+=0$; therefore, during numerical integration, we use $\ME_-=0$ in the region $r<r_0$ and $\ME_+=0$ in the region $r>r_0$ to obtain a solution connecting $a_S$ to $a_{den}$ (see Appendix~\ref{s:structure} for more details.).
To summarize Figs.~\ref{f:d_to_sch} --~\ref{f:h_to_sch},
for any value of $\beta$, we obtain a similar plot $a(r)$ like Fig.~\ref{f:uni_a}, which represents the simplest typical interior structure as Fig.~\ref{f:uni_pic}.
The central part is a flat region of size $r_0$; 
the middle part is a transient region of a higher energy density; 
the outermost part is described by the dense asymptotic metric \eqref{dense} 
with the near-Planckian curvature \eqref{R_den}; 
the surface is located at \eqref{surface}; 
it is connected to the exterior Schwarzschild metric.


As for the second task, we adopt Israel's junction condition~\cite{Poisson} at $r = r_{sur}$, and one can consult multishell models such as~\cite{KY2,KY3,KY4,Y1}\footnote{Regarding multishell models, we recommend readers consult, say, Sec.3 in~\cite{KY3} for gentle discussion and Sec.2 in~\cite{KY4} for details.} for concrete examples.
Here, we only mention the main idea.
That is, even though continuity of $a(r)$ seems insufficient to ensure the stability of spacetime, it is, in fact, not the case due to nontrivial surface tension at the junction.
In our case, the surface energy density $\epsilon_{sur}$ and surface pressure $p_{sur}$ are (see~\cite{KY2} for the calculation)
\begin{align}
    \epsilon_{sur} &= 0\\
    p_{sur} &= \frac{-1}{16\pi G}\sqrt{\frac{3}{2\gamma}}\left(2-\eta\right)
\end{align}
where the surface tension, $-p_{sur}$, is positive for~\eqref{eta}.


Note here that the proper length between $r_{sur}$ and $a_0$ is evaluated from \eqref{abc} and \eqref{dense} as 
$\sqrt{g_{rr}(R)}\f{2\g}{3\eta^2 a_0}=\MO(\sqrt{N}l_P)\gg l_P$ for $N\gg1$, 
which means that the quantum gravity effect is not large.
As a result of the semi-classical Einstein equation \eqref{Einstein}, the black hole has an interior dense region near the surface instead of a horizon~\cite{KMY,KY1,KY2,KY3,KY4,KY5,Y1,Ho_asymp}.

\section{Discussion: a conjecture on the origin of black hole entropy}\label{s:entropy}

Finally, we extend our result in Sec.~\ref{s:physical_structure} to discuss more general interior structures and then give a conjecture that the diversity of the interior structures is the origin of the Bekenstein-Hawking entropy~\cite{Bekenstein}.

\subsection{Diversity of the interior structure}\label{s:diversity}
Noting that a flat space can be seen as a Schwarzschild metric with zero mass, the simplest typical structure in Fig.~\ref{f:uni_pic} is a dense region sandwiched (approximatedly) by Schwarzschild metrics on both sides with different masses, and we call them interior and exterior Schwarzschild metrics. 
However, more complicated configurations satisfying the two physical conditions (no singularity and connection to the Schwarzschild metric with mass $a_0/2G$) can be obtained by using this typical structure.

\begin{figure}[!ht]
    \begin{minipage}[!ht]{0.45\linewidth}
        \centering
        \includegraphics[keepaspectratio, scale=0.5]{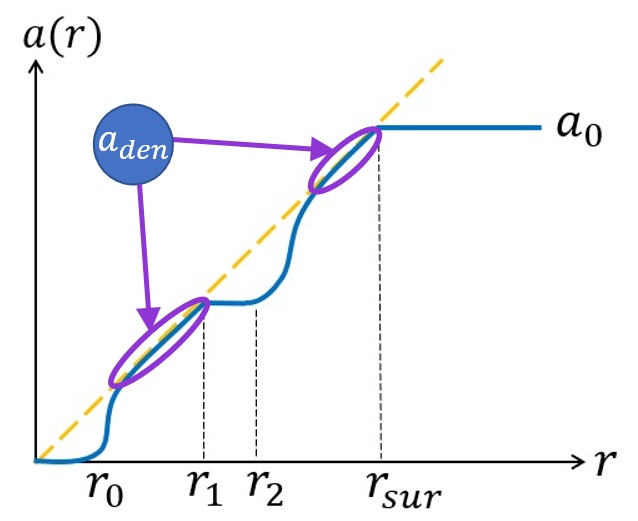}
        \caption{Plot of $a(r)$ for an example among the diverse interior configuration.}
        \label{f:div_a}
    \end{minipage}
    \hfill
    \begin{minipage}[!ht]{0.45\linewidth}
        \centering
        \includegraphics[keepaspectratio, scale=0.7]{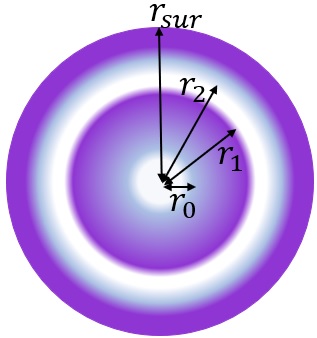}
        \caption{The interior configuration represented by Fig.~\ref{f:div_a}.}
        \label{f:div_pic}
    \end{minipage}
\end{figure}

A possible extension is to stack such a sandwich structure repeatedly, forming various configurations as in Fig.~\ref{f:div_a}.
This is possible because we are essentially matching the mass of the exterior Schwarzschild metric from one typical structure with that of the interior Schwarzschild metric from the other.
This is equivalent to inserting one dilute region, represented by the Schwarzschild metric, into the dense region of the typical structure, creating some hollow regions. 
(Note again here that a Schwarzschild metric is allowed as an approximated solution to the semi-classical Einstein equations \eqref{Einstein}, as discussed in Sec.~\ref{s:valid_eos}.)
For example, by inserting one dilute region, we obtain a configuration consisting of two concentric spherical shells of finite thickness (see Fig.~\ref{f:div_pic}.). 


Such a procedure for switching from the dense metric to the Schwarzschild metric 
is repeated several times to obtain a configuration consisting of several concentric spherical 
shells with finite thickness. 
For a given total mass $a_0/2G$, more shell structures can be created inside by reducing the switching interval. 
These various patterns of interior structures provide \textit{the diversity of the black hole interior}. 
Note again that identifying the dynamically stable one among them ultimately will require a time-dependent 
perturbation analysis using the semi-classical Einstein equations \eqref{Einstein}. 

\subsection{Conjecture toward entropy}
\label{s:conjecture}
This diversity leads naturally to the following conjecture: 
The possible patterns of classical geometry inside a black hole may explain the origin of black hole entropy. 
Roughly speaking, the two choices of whether it is the dense metric or the Schwarzschild-like one 
correspond to $\MO(1)$ bits of information per switching, 
and the number of possible patterns of interior structure may agree with the Bekenstein-Hawking formula.

\begin{figure}[!ht] 
    \centering
    \includegraphics[width=\textwidth]{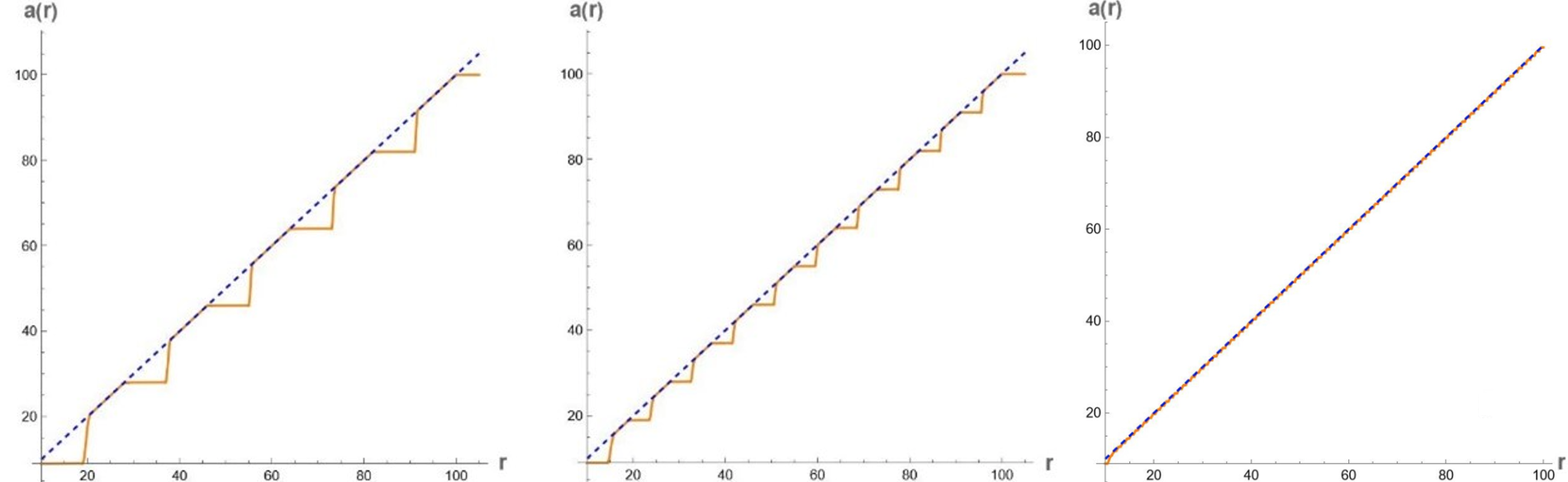}
    \caption{$a(r)$ ``oscillates" more finely for smaller switching intervals.}
    \label{f:switching}
\end{figure}

To investigate this, let us ask, 
``What is the minimum interval for switching between the high-density and low-density metrics?" 
This is determined by the equation of motion \eqref{eom}. 
For smaller switching intervals, $a(r)$ will oscillate more frequently, and gets increasingly closer to $a_{den}(r)$ (see Fig.~\ref{f:switching}.).
That indicates that for such a small interval, we can use the linearized equation of \eqref{eom} around $a_{den}(r)$ to understand the structure with the most number of switching, or equivalently, with the minimum interval for switching.

For detailed calculations of the linearized equation, one can refer to Appendix~\ref{A:f}.
Here, we only post the linearized solution $f_{den}(r)$ around $a_{den}(r)$ for any value of $\beta$:
\begin{align}\label{linear_sol}
    f_{den}(r) \in span \Big\{ &e^{-\lambda r^2} : \lambda=\frac{3\eta}{2\gamma}, \frac{3\eta^2}{4\gamma(2-\eta)}, \frac{3\eta}{8\gamma}\left(1\pm\sqrt{1+\frac{8\gamma}{3\beta}}\right) \Big\},
\end{align}
where the last one appears only for the cases of $\beta \neq 0$.
The plots in Fig.~\ref{f:switching} are repetitions of the flat part, the nearly vertical transition part, and the $a_{den}(r)$ part. 
The linearized solutions~\eqref{linear_sol} mean that the minimum width of the transient part around a point $r_*$ is determined dynamically as $\Delta r_{min} \equiv r-r_*\approx \frac{1}{\lambda r_*}\approx \frac{\gamma}{r_*}$, since
$e^{-\lambda r^2} = e^{-\lambda {r_*}^2} e^{-\lambda (r^2-{r_*}^2)}\approx e^{-\lambda {r_*}^2} e^{-2 \lambda r_*(r-{r_*})}$.
Converting this to the proper length $\Delta l_{min}=\sqrt{g_{rr}(r_*)} \Delta r_{min}$ (from the metric~\eqref{dense}), we obtain the minimum switching interval: 
\begin{equation}\label{mini}
    \D l_{min}\sim \sqrt{\gamma}\sim \sqrt{N}l_P, 
\end{equation}
which is the finest resolution of the interior structure.
Note that since this is a result obtained from the property of $a_{den}(r)$ in semi-classical gravity, the scale of $\D l_{min}$ is reliable\footnote{See Sec.~\ref{s:valid_eos} for a discussion on the validity of solutions.} with some large $N$.

Let us use this to evaluate the order of the entropy $S$. 
First, suppose we have a black hole of mass $\f{a_0}{2G}$ 
that contains only the switching structure for the minimum length \eqref{mini}. 
Then, the interior can be described approximately by the dense metric \eqref{dense}, 
which can be seen as the fact that the right-most figure in Fig.~\ref{f:switching} looks like $a(r)\approx a_{den}(r)$. 
The proper length of its size is given by 
\begin{equation}\label{l_size}
\begin{split}
    l_{size}
    =\int^{r_{sur}}_{r_0}dr\sqrt{g_{rr}(r)}
    \approx \sqrt{\f{3\eta^2}{8\g}}\left(r_{sur}^2-r_{0}^2\right)\sim\f{a_0^2}{\sqrt{N}l_P},        
\end{split}
\end{equation}
where \eqref{abc}, \eqref{dense}, and \eqref{surface} have been used. 
Therefore, by introducing $s_{switch}$, entropy per switching, the total entropy $S$ of the system should be expressed as
\begin{equation}\label{S_geo}
    S\sim s_{switch} \times \f{l_{size}}{\D l_{min}}\sim s_{switch} \f{a_0^2}{Nl_P{}^2}.
\end{equation}
First, \eqref{S_geo} means that, for a given total mass $\frac{a_0}{2G}$, the interior configuration with the largest number of switching will be statistically plausible. 
It can be approximated as a structure that oscillates slightly around the dense metric. 
Thus, a geometry like the one in Fig.~\ref{f:uni_pic} with as small $r_0$ as possible (maybe $r_0\approx \sqrt{N}l_P$ from \eqref{mini}) 
should be thermodynamically preferred. 
In short, configurations close to $a_{den}(r)$ with finest oscillations have the maximum entropy in the physical solutions of the equation of motion~\eqref{eom}, and, after smearing, it can be well-approximated by $a_{den}(r)$.

In order for \eqref{S_geo} to be consistent with the Bekenstein-Hawking entropy $S=\f{A}{4l_P{}^2}=\f{\pi a_0^2}{l_P{}^2}$, we must have 
\begin{equation}\label{s_sw}
    s_{switch}\sim N. 
\end{equation} 
Then, what is the origin of \eqref{s_sw}? 
Naively, $s_{switch}$ would be one bit of information corresponding to the two choices of the dense or dilute metrics. 
However, those metrics are the (approximate) solutions of the equation of motion \eqref{eom}, 
the semi-classical Einstein equation \eqref{Einstein} 
including the contribution from quantum matter fields with $\MO(N)$ degrees of freedom. 
Therefore, we can expect that 
with a switching, the configuration of the $\MO(N)$ matter fields also changes to match each metric, 
leading to \eqref{s_sw}.

Indeed, this is consistent with Bekenstein's idea \cite{Bekenstein}. 
Suppose we want to put a wave with one bit of information into a black hole of size $r$. 
To do so, we can use a wave with a wavelength $\sim r$. 
Since the wavelength and the size of the black hole are almost the same, 
the wave enters the black hole with a probability of about one-half 
and is bounced back with a probability of about one-half, 
which corresponds to about one bit of information. 
Such a wave has energy $\sim \frac{\hbar}{r}$. 
On the other hand, we can consider the black hole discussed above as consisting of many concentric spherical shells with widths in terms of the proper-length \eqref{mini}, and a shell at a point $r$ has the $r$-length width $\sim \f{\g}{r}$ and thus the energy $\sim \f{N\hbar}{r}$. 
Therefore, the shell contains $\MO(N)$ bits of information, which means \eqref{s_sw}. 
Thus, we have shown the possibility that the black-hole entropy comes from the number of possible patterns of classical interior geometry satisfying the semi-classical Einstein equation \eqref{Einstein}.

\section*{ACKNOWLEDGMENTS}
H.K. thanks Prof. Shin-Nan Yang and his family for their kind support through the Chin-Yu chair professorship. H.K. is partially supported by JSPS (Grants-in-Aid for Scientific Research Grants No.20K03970), by the Ministry of Science and Technology, R.O.C. (MOST 111-2811-M-002-016), and by National Taiwan University.
Y.Y. is partially supported by Japan Society of Promotion of Science 
(Grants No.21K13929 and No.18K13550) and by RIKEN iTHEMS Program. 

\appendix

\section{Examination of the equation of state}\label{A:eos}

To examine the equation of state for the interior region with a high density/curvature, for simplicity, we consider only the following ones:
\begin{align}
    \label{iso}\bra T^r{}_r \ket \propto \bra T^\theta{}_\theta \ket, \\
    \label{aniso}\bra -T^t{}_t \ket \propto \bra T^r{}_r \ket, \\
    \label{noniso}\bra -T^t{}_t \ket \propto \bra T^\theta{}_\theta \ket,
\end{align}
where the proportionality constants do not depend on spacetime.

\subsection{Why \texorpdfstring{\eqref{eos}}{}}\label{A:whyeos}

We consider such a dense configuration that $a(r) = r + \MO(r^0)$ is the leading order approximation for large $r$. 
Then, as we will see below, the solutions of \eqref{iso} and \eqref{noniso} are not thermodynamically stable. 
Therefore, we can conclude that \eqref{aniso}, or equivalently \eqref{eos}, is the only candidate equation of state for the dense interior. 

By the same procedure as in Sec.~\ref{s:master}, we use \eqref{iso} (instead of \eqref{eos}) and construct the self-consistent equation of $a(r)$. 
Then, we can choose the coefficient of \eqref{iso} properly and find the asymptotic solution starting from $a(r)=r$ in a similar method to Sec.~\ref{s:master}, as 

\begin{align}\label{quantities_iso}
\begin{split}
    a(r) &= r-\frac{16 \alpha ^2}{r^3} -\frac{512 \alpha ^3}{r^5} + O(r^{-7}),\\
    -G^t{}_t &= \frac{1}{r^2} + \frac{48 \alpha ^2}{r^6} + \frac{2560 \alpha ^3}{r^8} + O(r^{-10}),\\
    G^r{}_r &= -\frac{1}{r^2} - \frac{8 \alpha }{r^4}-\frac{144 \alpha ^2}{r^6} -\frac{128 \alpha ^2 (105 \alpha +8 \gamma )}{3 r^8} + O(r^{-10}),\\
    G^\theta{}_\theta &= \frac{1}{r^2} + \frac{8 \alpha }{r^4} + \frac{144 \alpha ^2}{r^6} + \frac{128 \alpha ^2 (105 \alpha +8 \gamma )}{3 r^8} + O(r^{-10}).
\end{split}
\end{align}
This has a negative radial pressure and is not stable thermodynamically.

For \eqref{noniso}, we can do a similar calculation to find
\begin{align}\label{quantities_noniso}
\begin{split}
    a(r) &= r-\frac{16 \alpha ^2}{r^3} +\frac{384 \alpha ^3}{r^5} + O(r^{-7}),\\
    -G^t{}_t &= \frac{1}{r^2} + \frac{48 \alpha ^2}{r^6} - \frac{1920 \alpha ^3}{r^8} + O(r^{-10}),\\
    G^r{}_r &= -\frac{1}{r^2} + \frac{8 \alpha }{r^4}-\frac{80 \alpha ^2}{r^6} + \frac{256 \alpha ^2 (33 \alpha +\gamma )}{3 r^8} + O(r^{-10}),\\
    G^\theta{}_\theta &= \frac{1}{r^2} + \frac{48 \alpha ^2}{r^6} -\frac{1920 \alpha ^3}{r^8} + O(r^{-10}).
\end{split}
\end{align}
In this case\footnote{Somehow \eqref{quantities_iso} and \eqref{quantities_noniso} are very similar. At leading order, each quantity is independent of $\g$, in contrast to the case with \eqref{aniso}, as shown in \eqref{G_den}.}, the radial pressure, again, is negative.

\subsection{Generality of (modified) Schwarzschild solution}\label{a:eos_sch}

We here see that modified Schwarzschid metrics appear in all cases of \eqref{iso}, \eqref{aniso}, and \eqref{noniso}.

With the equation of state \eqref{iso}, $\bra T^r{}_r \ket = w_1 \bra T^\theta{}_\theta \ket$, we can use the self-consistent equation for \eqref{iso} constructed when obtaining \eqref{quantities_iso}, to have the modified Schwarzschild metric as
\begin{align}
\begin{split}
    a(r) = a_0 + \frac{4 a_0^2 (\gamma -\alpha )}{r^3} -\frac{3 {a_0}^3 (w_1+2) (\alpha -\gamma )}{r^4 (10 w_1+4)} + O(r^{-5}).
\end{split}
\end{align}

Similarly, for the equation of state \eqref{noniso}, $\bra -T^t{}_t \ket = w_2 \bra T^\theta{}_\theta \ket$, we can find the modified Schwarzcshild metric as
\begin{align}
\begin{split}
    a(r) = a_0 + \frac{8 a_0^2 w_2 (\gamma -\alpha )}{r^3 \left(2 w_2-3\right)} -\frac{3 a_0^3 w_2 \left(2 w_2-1\right) (\alpha -\gamma )}{2 r^4 \left(2 w_2-3\right) \left(5 w_2-8\right)} + O(r^{-5}).
\end{split}
\end{align}

Since the Schwarzschild metric is always the leading behavior of the modified ones derived from various equations of state, we can argue that the ordinary Schwarzschild metric is a good approximation for regions with low curvatures.


\section{The full form of the self-consistent equation for \texorpdfstring{$a(r)$}{}}\label{A:eom}

We show the full form of the self-consistent equation for $a(r)$ \eqref{eom}, and for convenience, we write it again here:
\begin{align}
\beta a''''(r)
+\beta \frac{\left(2-\eta \right) \left(3-\eta \right) r a'(r)+\left(\eta +1\right) \eta a(r)-2 \left(\eta -1\right) \eta  r}{\left(2-\eta \right) \eta \left(r-a(r)\right) r}a'''(r)&\nonumber\\
+\frac{6 \beta -\gamma(2-\eta)}{3 \eta(r-a(r))} a''(r)^2+\mathcal{C}_1(r)a''(r)+\mathcal{C}_2(r)&=0,
\end{align}
where
\begin{align*}
\begin{split}
    \mathcal{C}_1(r)=&
    \frac{Q_1(r)}{3 (\eta -2) \eta ^2 (r-a(r))^2 r^2 }
    \\
    \mathcal{C}_2(r)=&
    \frac{3 \beta  \eta  (\eta +1)+\gamma  (\eta -2)}{3 \eta ^3 (r-a(r))^3} a'(r)^4
    +\frac{Q_2(r)}{3 (2-\eta ) \eta ^2 (r-a(r))^3 r} a'(r)^3 \\
    &+\frac{Q_3(r)}{3 (2-\eta) \eta  (r-a(r))^3 r^2}a'(r)^2
    -\frac{Q_4(r)}{(2-\eta) (r-a(r))^3 r^3}a'(r) \\
    &+\frac{12 \eta  (\alpha -\gamma )}{(2-\eta) r^4 (r-a(r))} a(r)^2,
\end{split}
\end{align*}
and
\begin{align*}
\begin{split}
    Q_1(r)=&
    (\eta -2) r^2 a'(r)^2 (\beta  (9 \eta +6)+2 \gamma  (\eta -2))+\eta  r a'(r) (3 \beta  (\eta  (\eta +2)-5) a(r)\\
    &+2 \gamma  (\eta -2) (4 \eta -5) a(r)-3 \beta  (\eta  (\eta +6)-8) r-8 \gamma  (\eta -2) (\eta -1) r) \\
    &+3 \eta ^2 a(r) \Big(a(r) (4 \alpha  (\eta -2)-5 \beta  \eta +\beta -4 \gamma  \eta +8 \gamma )\\
    &+r \left(-4 \alpha  \eta +8 \alpha +9 \beta  \eta -4
    \beta +4 \gamma  \eta -8 \gamma -\left((\eta -2) r^2\right)\right)\Big) \\
    &+3 \eta ^2 r^2 \left((\eta -2) r^2-4 \beta  (\eta -1)\right) 
\end{split}
\end{align*}
\begin{align*}
\begin{split}
    Q_2(r)=&
    r (3 \beta  (\eta +1) (3 \eta -4)+8 \gamma  (\eta -2) (\eta -1))-(\eta -2) a(r) (3 \beta +2 \gamma  (4 \eta -5))
\end{split}
\end{align*}
\begin{align*}
\begin{split}
    Q_3(r)=&
    r a(r) \Big(2 \eta  (6 \alpha  (5-2 \eta )+3 \beta  (4 \eta -5)+2 \gamma  (8 \eta -21))\\
    &-24 \alpha +3 \beta +64 \gamma +3 (\eta -2) r^2\Big)\\
    &+a(r)^2 (12 \alpha  (\eta -2) (\eta -1)+3 \beta  (5-4 \eta ) \eta +\gamma  (4 (13-4 \eta ) \eta -49))\\
    &+r^2 \left(6 \eta  (2 \alpha  (\eta -2)-2 \beta  \eta +\beta )-16 \gamma  (\eta -1)^2-3 (\eta -2) r^2\right)
\end{split}
\end{align*}
\begin{align*}
\begin{split}
    Q_4(r)=&
    a(r) \Big(r a(r) \left(4 \eta  (-8 \alpha +5 \beta +8 \gamma )+36 \alpha -11 \beta -36 \gamma +(2 \eta -1) r^2\right)\\
    &+4 a(r)^2 (\alpha  (4 \eta -5)-2 \eta  (\beta +2 \gamma
    )+\beta +5 \gamma )+(3-4 \eta ) r^4\\
    &+2 r^2 (8 \alpha  (\eta -1)-8 \eta  (\beta +\gamma )+5 \beta +8 \gamma )\Big)+2 (\eta -1) r^3 \left(2 \beta +r^2\right)
\end{split}
\end{align*}


\section{Asymptotic solution due to single term cancellation}\label{A:k}

In this appendix, we are going to prove the statement mentioned in Sec.~\ref{s:sol}: single leading term cancellation can happen only when $k > 3$. As shown in Fig.~\ref{f:k_plot}, we have a single leading term in three possible domains: $k<1$, $1<k<3$, and $k>3$. Therefore, to prove the statement, it is sufficient to consider the first two cases.

When $k<1$, the leading term with its coefficient is
\begin{align*}
\begin{split}
    -3 \eta ^3 k (\eta -(2-\eta) k) r^{k+5},
\end{split}
\end{align*}
and its coefficient vanishes when $k=0, \frac{\eta}{2-\eta}$. 
When $k=0$, it is covered in $a_S$. 
As for $k= \frac{\eta}{2-\eta}$, it can only take value at least $1$ within $1\leq \eta \leq 2$. 
Hence, we do not generate new cases here.

Next, we have a look at $1<k<3$. The leading term with its coefficient is given by
\begin{align*}
\begin{split}
    -3 \eta ^2 k \left(\eta (\eta +1) +(\eta ^2-3 \eta +2) k\right) r^{3k+3},
\end{split}
\end{align*}
and its coefficient vanishes when $k=0, -\frac{\eta  (\eta +1)}{\eta ^2-3 \eta +2}$. $k=0$ is not in the range of $k$ here. For the other one, $-\frac{\eta  (\eta +1)}{\eta ^2-3 \eta +2}$, it has minimum at $\eta=\frac{1}{2} \left(1+\sqrt{3}\right)$ with value $k=7+4 \sqrt{3} >3$, so again, it is a contradiction.

To summarize, we see that beside the asymptotic solutions mentioned in Sec.~\ref{s:sol}, the only possibilities are those with leading order $r^k$ where $k > 3$. However, we can check numerically that such solutions go quickly to singularities as $r$ increases, and we do not include them in our analysis.


\section{Full result of series expansion and their physical properties}\label{A:full}

In this appendix, we are going to explain in detail about the solution series mentioned in Sec.~\ref{s:sol}.
The solutions we focus on here are those with leading order $a(r)\sim \MO(C+r^{-3}),~\MO(r),~\MO(r^3)$.

First, we demonstrate explicitly how to construct it for $a(r)=\MO(C+r^{-3})$.
Setting 
\begin{equation}
a(r)=c_0+\f{c_1}{r^3}+\f{c_2}{r^4}+\f{c_3}{r^5}+\cdots,
\end{equation}
and expanding \eqref{eom} for $r\gg l_P$, 
we obtain 
\begin{align}
&\frac{12 {c_0}^2 (\alpha -\gamma )-\frac{6 {c_1} (\eta -3)}{\eta }}{r^6}+\frac{4 {c_2} (8-3 \eta )-3 {c_0} {c_1}}{\eta  r^7}\nonumber\\
&~~~~~~~~~~~~~~~~~~~+\frac{-3 {c_0}^2 {c_1}-4 {c_0} {c_2}-180 \beta  {c_1} (\eta -3)+10 {c_3} (5-2 \eta )}{\eta  r^8}+\MO(r^{-9})=0.
\end{align}
From the first term, we get $c_1=\frac{2 {c_0}^2 \eta  (\alpha -\gamma )}{\eta -3}$; 
from the second one, we find ${c_2}=\frac{3 {c_0} {c_1}}{32-12 \eta }$;
from the third one, we obtain $c_3=-\frac{3 {c_0}^2 {c_1}+4 {c_0} {c_2}+180 \beta  {c_1} (\eta -3)}{20 \eta -50}$, and so on. 
Thus, we can determine each coefficient $c_i$ order by order in terms of a given constant $c_0$ 
to reach the series solution $a(r)=a_S(r)$, where  
\begin{equation}
\label{a_Sa}
a_S(r)=c_0 +\frac{2 {c_0}^2 \eta  (\alpha -\gamma )}{(\eta -3) r^3}
+\frac{3 {c_0}^3 \eta  (\alpha -\gamma )}{2(\eta -3)(8-3 \eta ) r^4}
+\frac{9 {c_0}^2 \eta  (\alpha -\gamma ) \left(-20 \beta  (8-3 \eta)+{c_0}^2\right)}{5 (2 \eta -5) (8-3 \eta) r^5}+\MO(r^{-6}),
\end{equation}
which is a deformation of the Schwarzschild metric of mass $\f{c_0}{2G}$. 

Similarly, we can find the series solutions for the other cases. 
For $a(r)=\MO(r)$, we have 
\begin{align}\label{a_eta}
&a_\eta(r) =\frac{\eta  (\eta -1) }{\eta ^2-\eta +1} r
+ \frac{2 \eta  (\eta -1)^2 (2 \alpha  (\eta -2) \eta +3 \gamma )}{\left(\eta ^2-\eta +1\right)^2 \left(\eta ^3-\eta ^2-2 \eta +3\right) r}\nonumber\\
&-\f{4 (\eta -1)^2 \eta  (2 \alpha  (\eta -2) \eta +3 \gamma )}{\left(\eta ^2-\eta +1\right)^3 \left(\eta ^3-10 \eta +15\right) \left(\eta ^3-\eta ^2-2 \eta +3\right)^2 r^3}
\l[2 \beta  (4-\eta) \left(\eta ^3-\eta ^2-2 \eta +3\right)^2\r.\nonumber\\
&\l.+(\eta -1) \left(\gamma  \left(2 \eta ^6-\eta ^5-24 \eta ^4+35 \eta ^3+23 \eta ^2-72 \eta +33\right)-2 \alpha  \eta  \left(2 \eta ^5-8 \eta ^4+5 \eta ^3+10 \eta ^2-13 \eta +1\right)\right)\r]\nonumber\\
&+\MO(r^{-5}),
\end{align}
where the slope of the leading term depends on the value of $\eta$. 

For $a(r)=\MO(r)$, there is the other possibility: $a(r)=r+\cdots$. 
This is because the term $r-a(r)$ appears in the denominator of \eqref{eom}, which requires a special treatment.
Then, we have 
\begin{align}
\label{a_dena}
&a_{den}(r)=r-\frac{2 \gamma }{3 \eta ^2 r}+\frac{4 \gamma  \left(\gamma  \left(\eta ^2+6 \eta -2\right)-6 \alpha  \eta ^2\right)}{9 \eta ^4 r^3}\nonumber\\
&-\frac{8 \gamma  \left(-6 \gamma  \eta ^2 \left(\alpha  \left(\eta ^2+15 \eta -7\right)-3 \beta \right)+18 \alpha  \eta ^3 (2 \alpha  \eta -\beta )+\gamma ^2 \left(\eta ^4+13 \eta ^3+55 \eta ^2-58 \eta +13\right)\right)}{27 \eta ^6 r^5} \nonumber \\
&+\MO(r^{-7}),
\end{align}
which represents a dense interior in the sense that $a_{den}(r)\approx r$. 

We finally obtain for $a(r)=\MO(r^3)$ 
\begin{align}\label{a_neg}
&a_{neg}(r) =\frac{\eta ^2 \left(2 \eta ^2-4 \eta +3\right) r^3}{2 (\eta -3) \left(2 \alpha  (\eta -1) \eta ^2-\gamma  (\eta -3)\right)} \nonumber\\
&+ \frac{3 r \left(2 \alpha  \eta ^4+\beta  \left(-2 \eta ^3+10 \eta ^2-15 \eta +9\right) \eta +\gamma  \left(4 \eta ^3-19 \eta ^2+24 \eta -9\right)\right)}{\eta  \left(2 \alpha  \eta  \left(\eta ^3-8 \eta ^2+15 \eta -9\right)+\beta  \left(-2 \eta ^4+2 \eta ^3+7 \eta ^2-15 \eta +9\right)\right)+\gamma  \left(4 \eta ^4+9 \eta ^3-40 \eta ^2+42 \eta -9\right)} \nonumber\\
&+\MO(r^{-1}),
\end{align}
which has a negative energy, as shown in Sec.~\ref{s:prop}. 

Note that for all the cases, the effects of $\b$ are involved only from subleading terms. 

Here, we check which of these four solutions belongs to the $\ME_+$ branch or the $\ME_-$ branch.
Setting $\b=0$ and substituting these into \eqref{eom0}, 
we can see that 
$a_S(r)$, $a_\eta(r)$, and $a_{neg}(r)$ satisfy only $\ME_-=0$, 
while $a_{den}(r)$ meets only $\ME_+=0$. 
Therefore, $a_S(r)$, $a_\eta(r)$, and $a_{neg}(r)$ belong to the $\ME_-$ branch, 
while $a_{den}(r)$ belongs to the $\ME_+$ one. 

\subsection{Physical properties }\label{s:prop}
Now, we study the physical properties of the four asymptotic solutions. 
\subsubsection*{Dilute density metric $a_S(r)$}
From \eqref{metric} with \eqref{A}, the leading behavior of the metric for \eqref{a_Sa} becomes 
\begin{equation}
\label{dilutea}
ds^2\approx-\l(1-\f{c_0}{r}+\frac{2 \eta  (\alpha-\g ){c_0}^2}{(3-\eta) r^4}\r)
e^{-\frac{3\eta  (\alpha-\g ){c_0}^2}{(3-\eta) r^4}}dt^2
+\l(1-\f{c_0}{r}+\frac{2\eta  (\alpha-\g ){c_0}^2}{(3-\eta) r^4}\r)^{-1}dr^2+r^2d\Omega^2,
\end{equation}
where we have absorbed the integration constant from (2.6) to the redefinition of $t$ and kept the leading correction.
This metric is controlled by a length-dimensional constant $c_0$. 
Also, it is perturbative in $\hbar$ since it is still finite as $\hbar\rightarrow 0$.

We can calculate the Einstein tensors as
\begin{equation}
-G^t{}_t=\frac{6 \eta  (\alpha-\g ){c_0}^2}{(3-\eta) r^6},~
G^r{}_r=\frac{6 (2-\eta)(\alpha-\g ){c_0}^2}{(3-\eta) r^6},~
G^\theta{}_\theta=-\frac{12 (2-\eta) (\alpha-\g ){c_0}^2}{(3-\eta) r^6},
\end{equation}
with the sub-leading terms of $\MO(r^{-7})$, and the curvatures as 
\begin{align}
R=\frac{12 (\alpha -\gamma )c_0^2}{r^6}+\MO(r^{-8}),&~~
R_{\mu\nu\a\b}R^{\mu\nu\a\b}
=\frac{12 {c_0}^2}{r^6}+\frac{48 (6-5 \eta) (\alpha -\gamma )c_0^3}{(3-\eta) r^9}+\MO(r^{-10}),\nonumber\\
R_{\mu\nu}R^{\mu\nu}&=\frac{72 \left(5 \eta ^2-18 \eta +18\right) (\alpha -\gamma )^2c_0^4}{(3-\eta)^2 r^{12}}+\MO(r^{-13}).
\end{align}
For \eqref{eta} and \eqref{a/c}, the energy density $\bra -T^t{}_t\ket=\f{1}{8\pi G} (-G^t{}_t)$ 
can be positive and negative, depending on the ratio $\a/\g$. 
For $c_0\sim r \gg l_P$ (at most), 
the energy density and curvatures are small compared to $\MO(1)$ (from \eqref{abc}). 
Therefore, \eqref{dilutea} represents a modified Schwarzschild metric with small quantum corrections.

\subsubsection*{Moderate density metric $a_\eta(r)$}
The leading behavior of the metric for \eqref{a_eta} is given by 
\begin{equation}\label{moderate}
ds^2\approx-\frac{1}{\eta ^2-\eta +1}\l(\f{r}{r_0}\r)^{2(\eta-1)}dt^2+(\eta ^2-\eta +1)dr^2+r^2d\Omega^2,
\end{equation}
which is parametrized only by $\eta$, and $g_{rr}(r)$ is positive and constant for \eqref{eta}. 
Note that the metric for $a_\eta(r)$ has a $\hbar \rightarrow 0$ limit (see Appendix \ref{A:eta}.).
This gives 
\begin{equation}\label{G_eta}
-G^t{}_t=\frac{(\eta -1) \eta }{\left(\eta ^2-\eta +1\right) r^2},~
G^r{}_r=-\frac{(2-\eta) (\eta -1)}{\left(\eta ^2-\eta +1\right) r^2},~
G^\theta{}_\theta=\frac{(\eta -1)^2}{\left(\eta ^2-\eta +1\right) r^2},
\end{equation}
with the sub-leading terms of $\MO(r^{-4})$, and 
\begin{align}\label{mod_R}
R=0
+\MO(r^{-4}),&~~
R_{\mu\nu\a\b}R^{\mu\nu\a\b}
=\frac{8 (\eta -1)^2 \left(\eta ^2-2 \eta +3\right)}{\left(\eta ^2-\eta +1\right)^2 r^4}+\MO(r^{-6}),\nonumber\\
R_{\mu\nu}R^{\mu\nu}&=\frac{2 (\eta -1)^2 \left(2 \eta ^2-4 \eta +3\right)}{\left(\eta ^2-\eta +1\right)^2 r^4}+\MO(r^{-6}).
\end{align}

\begin{figure}[!ht] 
    \centering
    \includegraphics[scale =0.7]{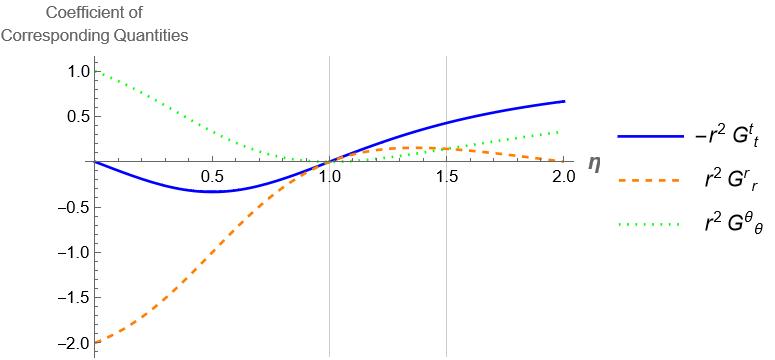}
    \caption{The plot of \eqref{G_eta} for $0<\eta<2$.
    The blue (solid), orange (dashed), and green (dotted) lines represent $-r^2G^t{}_t$, $r^2G^r{}_r$, and $r^2G^\theta{}_\theta$, respectively.}
    \label{f:eta}
\end{figure}

The energy density and the curvatures are in the order of $\MO(r^{-2})$, 
which is small compared to $\MO(1)$ but larger than those of $a_S(r)$. 
For \eqref{eta}, each of \eqref{G_eta} behaves like Fig.~\ref{f:eta}. 
Note here that for $1\leq \eta<2$ the dominant energy condition is satisfied as a normal matter. 
Each component of $G^\mu{}_\nu$ is $\MO(r^{-2})$ while the Ricci scalar is $R=0+\MO(r^{-4})$, 
which means that the leading power in $G^\mu{}_\mu$ cancels,\footnote{
More precisely, 
we have 
$R=-\frac{4 \left((\eta -1)^2 (-2 \alpha  (2-\eta) \eta +3 \gamma )\right)}{\left(\eta ^2-\eta +1\right)^2 r^4}+\MO(r^{-6})$, 
and both sides of \eqref{eom} are balanced at the sub-leading order for any $\a$ and $\g$.} 
and the leading metric \eqref{moderate} represents a conformal matter for $1\leq \eta<2$.
Indeed, for $\eta=\f{3}{2}$, \eqref{moderate} reproduces the metric of conformal fluid.\footnote{
The isotropic condition $G^r{}_r=G^\theta{}_\theta$ means $\eta=\f{3}{2}$ (from Fig.~\ref{f:eta}). 
Then, \eqref{G_eta} gives $-G^t{}_t=\f{3}{7r^2}$ and $G^r{}_r=G^\theta{}_\theta=\f{1}{7r^2}$, 
and \eqref{moderate} becomes 
\begin{equation}
ds^2=-\f{4}{7}\f{r}{r_0}dt^2+\f{7}{4}dr^2+r^2 d\Omega^2,\nonumber
\end{equation}
which agrees with the metric of a spherical static conformal fluid \cite{Weinberg}.}$^{,}$\footnote{Note that for $\eta=1$, \eqref{moderate} becomes the flat metric, and \eqref{G_eta} and \eqref{mod_R} vanish.} 


Thus, the metric for $a_\eta(r)$ has moderate energy density and curvatures, and expresses the anisotropic configuration characterized by $\eta\in [1,2)$ of a conformal matter, including the quantum corrections. 
Note that the metric depends only on $(\eta;\g,\a,\b)$, not on a given scale such as $c_0$ in \eqref{dilutea}. 

\subsubsection*{High density metric $a_{den}(r)$}
The leading behavior of the metric for \eqref{a_dena} is 
\begin{equation}
\label{densea}
ds^2\approx-\frac{2 \gamma }{3 \eta ^2 r^2}e^{\frac{3 \eta  r^2}{2 \gamma }}dt^2+\frac{3 \eta ^2 r^2}{2 \gamma }dr^2+r^2 d\Omega^2,
\end{equation}
which depends only on $\g$ and $\eta$. This has 
\begin{equation}
-G^t{}_t=\f{1}{r^2}+\MO(r^{-4}),~
G^r{}_r=\frac{2-\eta}{\eta}\f{1}{r^2}+\MO(r^{-4}),~
G^\theta{}_\theta=\frac{3}{2\g}+\MO(r^{-2})
\end{equation}
and 
\begin{equation}
\label{R_dena}
R=-\f{3}{\g},~~R_{\mu\nu\a\b}R^{\mu\nu\a\b}=\f{9}{\g^2},~~R_{\mu\nu}R^{\mu\nu}=\f{9}{2\g^2},
\end{equation}
with the sub-leading corrections of $\MO(r^{-2})$. 

First, the metric is non-perturbative for $\hbar$, since we cannot take the limit $\hbar\to0$. 
Second, it is uniform and almost Planckian in that 
the curvature invariants are constants of $\MO\l(\f{1}{N l_P{}^2}\r)$, 
which is still smaller than the Planck scale for a large $N$. 
In this sense, this represents a semi-classical, dense configuration.
Third, from \eqref{R_dena}, we have $R=-\f{2}{L^2}+\MO(r^{-2})$, 
which indicates that the metric is a warped product of $AdS_2$ of radius $L\equiv\sqrt{\f{2\g}{3}}\sim \sqrt{N}l_P$ 
and $S^2$ of radius $r$.\footnote{\label{foot:AdS}More explicitly, we can check it as follows \cite{KY4,KY5}. 
In general, $ds^2=-f(r)dt^2+h(r)dr^2+r^2d\Omega^2=\f{L^2}{z^2}(-dt^2+dz^2)+r(z)^2d\Omega^2$ 
holds if the condition $\sqrt{h(r)}=\pm\f{L}{2}\p_r \log f(r)$ is satisfied. 
Using \eqref{densea} and focusing a point $r$, 
we have $h(r)=\frac{3 \eta ^2 r^2}{2 \gamma }$ and $\p_r \log f(r)\approx \p_r A(r)=\frac{3 \eta  r}{\gamma }$, 
and then the condition holds for $L=\sqrt{\f{2\g}{3}}$.} 
Fourth, the tangential pressure $\bra T^\theta{}_\theta\ket$ is so large that $-G^t{}_t,G^r{}_r \ll G^\theta{}_\theta$, breaking the dominant energy condition. 
This comes from the vacuum fluctuations of all the modes with any angular momentum \cite{KY4}. 
Fifth, the energy density and pressures are positive for \eqref{eta}, 
and thus this metric can represent a physical excitation.


\subsubsection*{Negative energy metric $a_{neg}(r)$}
\begin{figure}[!ht] 
\centering
\includegraphics[scale =0.3]{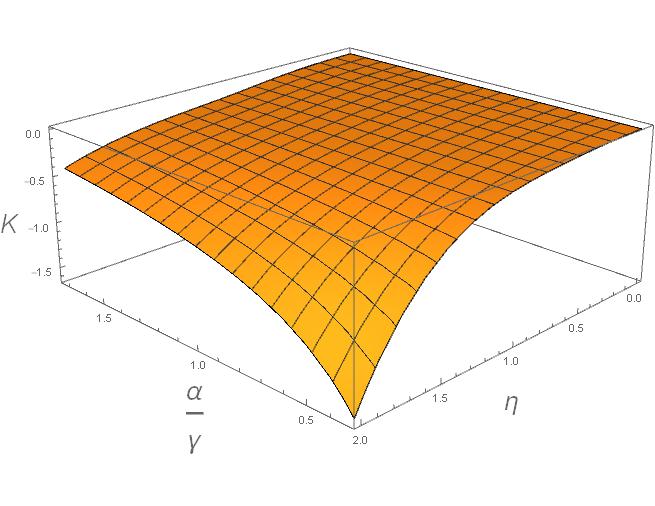}
\caption{The value of $K$ \eqref{K} for the range of \eqref{eta} and \eqref{a/c}.}
\label{f:K_nega}
\end{figure}

Finally, we study the leading behavior of the metric for \eqref{a_neg}: 
\begin{align}\label{negative}
ds^2&\approx-\l(1-K r^2\r)\l(\f{r_0}{r}\r)^{\f{6}{\eta}}dt^2+ \l(1-Kr^2\r)^{-1}dr^2+r^2 d\Omega^2,
\end{align}
where 
\begin{equation}\label{K}
K\equiv \frac{\eta ^2 (2 (2-\eta) \eta -3)}{2 (3-\eta) \left(2 \alpha  (\eta -1) \eta ^2+\gamma  (3-\eta)\right)}
\end{equation}
is of the order $\MO\l(\f{1}{Nl_P{}^2}\r)$ and negative for \eqref{eta} and \eqref{a/c} (see Fig.~\ref{f:K_nega}.).

This leads to, up to the sub-leading corrections of $\MO(r^{-2})$, 
\begin{align}
-G^t{}_t=&-\frac{3 \eta ^2 \left(2 \eta ^2-4 \eta +3\right)}{2 (3-\eta) \left(2 \alpha  (\eta -1) \eta ^2+\gamma  (3-\eta)\right)},~~
G^r{}_r=-\frac{3 \eta(2-\eta) \left(2 \eta ^2-4 \eta +3\right)}{2 (3-\eta) \left(2 \alpha  (\eta -1) \eta ^2+\gamma  (3-\eta)\right)},\nonumber\\
&~~~~~~~~~~~~~~G^\theta{}_\theta=\frac{3 \left(\left(\eta ^2-3 \eta +3\right) \left(2 \eta ^2-4 \eta +3\right)\right)}{2 \left((3-\eta) \left(2 \alpha  (\eta -1) \eta ^2+\gamma  (3-\eta)\right)\right)}
\end{align}
and
\begin{align}
R &= - \frac{3 \left(2 \eta ^2-4 \eta +3\right)^2}{(3-\eta) \left(2 \alpha  (\eta -1) \eta ^2+\gamma  (3-\eta)\right)},~
R_{\mu\nu\a\b}R^{\mu\nu\a\b}
=\frac{3 \left(2 \eta ^2-4 \eta +3\right)^2 \left(2 \eta ^4-8 \eta ^3+24 \eta ^2-36 \eta +27\right)}{(3-\eta)^2 \left(\gamma  (3-\eta)+2 \alpha  (\eta -1) \eta ^2\right)^2},\nonumber\\
&~~~~~~~~~~~~~~~~~~~~
R_{\mu\nu}R^{\mu\nu}= \frac{9 \left(\eta ^2-2 \eta +3\right) \left(2 \eta ^2-4 \eta +3\right)^3}{2 (3-\eta)^2 \left(\gamma  (3-\eta)+2 \alpha  (\eta -1) \eta ^2\right)^2}.
\end{align}

This metric is also non-perturbative for $\hbar$ and near Planckian. 
The most important point is that, from \eqref{Mr} and \eqref{K}, it has a large negative energy for $r\gg l_P$: 
\begin{equation}
\label{neg_energy}
M(r)=\f{a_{neg}(r)}{2G}= \f{Kr^3}{2G}<0. 
\end{equation}
The magnitude $|M(r)|\sim \f{r^2}{Nl_P{}^2}\f{r}{G}$ is too large for a black hole with mass $\sim \f{r}{2G}$. 
Therefore, this metric cannot be considered a physical metric for the interior, and we don't consider it in the following.


\section{On moderate density metric \texorpdfstring{$a_\eta(r)$}{}}\label{A:eta}

We here calculate the subleading terms to \eqref{moderate}, \eqref{G_eta}, and \eqref{mod_R} to see that the geometry of $a_\eta(r)$ is perturbative for $\hbar$.

The metric for \eqref{a_eta} is given by 
\begin{align}
\begin{split}
    ds^2
    \approx
    &-\left(\eta ^2-\eta +1 + \frac{2 (\eta -1)^2 \eta  (2 \alpha  (\eta -2) \eta +3 \gamma )}{((\eta -2) \eta  (\eta +1)+3) r^2}\right)^{-1} 
    R(r) dt^2\\
    &+\left(\eta ^2-\eta +1 + \frac{2 (\eta -1)^2 \eta  (2 \alpha  (\eta -2) \eta +3 \gamma )}{((\eta -2) \eta  (\eta +1)+3) r^2}\right)dr^2 \\
    &+r^2d\Omega^2,
\end{split}
\end{align}
where
\begin{align}
    R(r)=r^{2 \eta -4} \left(\left(\eta  \left(\eta ^4-2 \eta ^3+4 \eta -5\right)+3\right) r^2-2 (\eta -1)^2 ((\eta -1) \eta -1) (2 \alpha  (\eta -2)
   \eta +3 \gamma )\right).
\end{align}
This gives 
\begin{align}
\begin{split}
    -G^t{}_t &=\frac{(\eta -1) \eta}{(\eta^2 -\eta +1)r^2} + \frac{2 (\eta -1)^2 \eta  (2 \alpha  (\eta -2) \eta +3 \gamma )}{((\eta^2 -\eta +1)^2 ((\eta -2) \eta  (\eta +1)+3) r^4}+\MO(r^{-6})\\
    G^r{}_r &=-\frac{(\eta -2) (\eta -1)}{(\eta^2 -\eta +1) r^2}+\frac{2 (\eta -2) (\eta -1)^2 (2 \alpha  (\eta -2) \eta +3 \gamma )}{(\eta^2 -\eta +1)^2 ((\eta -2) \eta  (\eta +1)+3) r^4}+\MO(r^{-6}), \\
    G^\theta{}_\theta &=\frac{(\eta -1)^2}{(\eta^2 -\eta +1)r^2} + \frac{2 (\eta -1)^2 (\eta  ((\eta -1) \eta -3)+4) (2 \alpha  (\eta -2) \eta +3 \gamma )}{(\eta^2 -\eta +1)^2 ((\eta -2) \eta  (\eta +1)+3)
   r^4}+\MO(r^{-6}),
\end{split}    
\end{align}
and 
\begin{align}
\begin{split}
    R=&-\frac{4 \left((\eta -1)^2 (-2 \alpha  (2-\eta) \eta +3 \gamma )\right)}{\left(\eta ^2-\eta +1\right)^2 r^4} +\MO(r^{-6}),\\ 
    R_{\mu\nu\a\b}R^{\mu\nu\a\b}=&\frac{8 (\eta -1)^2 \left(\eta ^2-2 \eta +3\right)}{\left(\eta ^2-\eta +1\right)^2 r^4}\\
    &+\frac{16 (\eta -1)^3 (\eta  (\eta  ((\eta -3) \eta -2)+17)-12) (2 \alpha  (\eta -2) \eta +3 \gamma )}{((\eta -1) \eta +1)^3 ((\eta -2) \eta 
   (\eta +1)+3) r^6}+\MO(r^{-8}),\\
    R_{\mu\nu}R^{\mu\nu}=&\frac{2 (\eta -1)^2 \left(2 \eta ^2-4 \eta +3\right)}{\left(\eta ^2-\eta +1\right)^2 r^4}\\
    &+\frac{8 (\eta -2) (\eta -1)^3 \left(\eta ^3-3 \eta +3\right) (2 \alpha  (\eta -2) \eta +3 \gamma )}{((\eta -1) \eta +1)^3 ((\eta -2) \eta 
   (\eta +1)+3) r^6}+\MO(r^{-8}).    
\end{split}
\end{align}


\section{Structure of the solution space}\label{s:structure}

In this appendix, we examine the general features of the solutions based on the asymptotic solutions in two steps. 

First, in Sec.~\ref{s:local}, we consider the local structure of the solution space around each asymptotic solution. That is, we analytically solve the linearized equation around each asymptotic solution and see how the solutions deviate from it. 

These deviations are said to be relevant/irrelevant if they increase/decrease as $r$ increases.
More precisely, we are looking at 
\begin{align}
    a(r) = a_{asy}(r) \left( 1 + \epsilon \mathcal{D}(r) \right),
\end{align}
where $\e$ is an infinitesimal parameter, and $\mathcal{D}(r)$ represents a deviation from $a_{asy}(r)$. 
A deviation, $\mathcal{D}(r)$, is defined as relevant/irrelevant if it grows/decays as $r$ increases.
In other words, the relevant/irrelevant deviations lead us away/toward the corresponding asymptotic solution.
In this sense, we call them relevant/irrelevant directions of an asymptotic solution, where $r$ is assumed to increase. 

Note that there are also deviations, $\mathcal{D}(r)$'s, whose absolute values approach constant in the large $r$ region.
We call them marginal directions, and their relevancy should be determined by the perturbation of the next order. In our case, we check them numerically to identify whether they are marginally relevant or marginally irrelevant.

Then, one would expect that the general solution can be viewed as a sequence of transitions from one asymptotic solution to another.
To see this, we need to consider not only small deviations from the asymptotic solutions but also finite deviations. 
This will be checked numerically in Sec.~\ref{s:global}. 
In this way, the global structure of the solution space is obtained.




\subsection{Local behaviors around the asymptotic solutions}\label{s:local}
First, we consider perturbation analysis in the neighborhood of each asymptotic solution to identify the relevant and irrelevant directions.

We construct the linearized equation for the self-consistent equation for $a(r)$ \eqref{eom} 
around each asymptotic solution $a_{asy}(r)$, where we consider only the leading terms for large $r$.
Setting\footnote{Note that, in the following, we define $f(r) = a_{asy}(r) \mathcal{D}(r)$, and consider equations in the form of $f(r)$'s (see \eqref{e:deviation}.), since this simplifies the form of equations. However, we shall keep in mind that we should focus on $\mathcal{D}(r)$ when we want to identify whether a deviation (or direction) is relevant or irrelevant.}
\begin{equation}\label{e:deviation}
    a(r)=a_{asy}(r)+\e f(r)
\end{equation}
with a small parameter $\e$, expanding \eqref{eom} to $\MO(\e)$, and picking up the leading terms for $r\gg l_P$, 
we can obtain, for $a_S(r)$ \eqref{a_Sa}, $a_\eta(r)$ \eqref{a_eta}, and $a_{den}(r)$ \eqref{a_dena}, respectively, 
\begin{align}\label{f_S}
  \b f_{S}''''(r)
    &-\frac{2 \b(\eta -1) }{(2-\eta ) r} f_{S}'''(r)
    +f_{S}''(r)
    -\frac{2 (\eta-1)}{(2-\eta ) r}f_{S}'(r) = 0 
    \\
    \label{f_eta}
    \b f_{\eta}''''(r)
    &+\frac{2 \b (\eta -1) }{r}f_{\eta}'''(r)
    +\left(\eta ^2-\eta+1\right)f_{\eta}''(r) \nonumber\\
    &+\frac{(\eta -1) \left(\eta ^2-\eta+1\right)}{r}f_{\eta}'(r)
    +\frac{\eta (\eta -1)(2 \eta -1)\left(\eta ^2-\eta +1\right) }{(2-\eta ) r^2}f_{\eta}(r) = 0
    \\
    \label{f_den}
    \b \bar f_{den}''''\left(r^2\right)
    &+\frac{3 \b (3-\eta ) \eta}{2 \gamma  (2-\eta)}\bar f_{den}'''\left(r^2\right)
    -\frac{3 \eta ^2 (2\gamma  (2-\eta )-3 \beta  (\eta +4))}{16\gamma^2 (2-\eta)}\bar f_{den}''\left(r^2\right)\nonumber\\
    &-\frac{9 \eta ^3 (\gamma  (4-\eta )-3\beta \eta)}{32 \gamma ^3 (2-\eta)}\bar f_{den}'\left(r^2\right)
    -\frac{27 \eta ^5}{64 \gamma ^3(2-\eta)}\bar f_{den}\left(r^2\right) = 0.
\end{align}
Here, to get the last one, we have set $f_{den}(r)=\bar f_{den}(r^2)$ in the linearized equation, expanded it again, and picked up only the leading terms for $r\gg l_P$. 

We can solve these to get (See Appendix \ref{A:f} for the derivations.) 
\begin{align}
\label{f_S+}
    f_{S}^{(\b>0)}(r) \in span \Big\{ &1, r^{\frac{\eta }{2-\eta }}, e^{\pm i \f{r}{\sqrt{\b}}} \Big\}, \\
\label{f_S-}
    f_{S}^{(\b<0)}(r) \in span \Big\{ &1, r^{\frac{\eta }{2-\eta }}, e^{\pm \f{r}{\sqrt{-\b}}} \Big\}, \\
\label{f_eta+}
    f_{\eta}^{(\b>0)}(r) \in span \Big\{& r^{\frac{1}{2} \left(2-\eta\pm\sqrt{9 \eta ^2+\frac{24}{\eta -2}+16}\right)}, 
    r^{\frac{1-\eta }{2}} e^{\pm i r\sqrt{\frac{(\eta -1) \eta +1}{\beta }}} \Big\}, \\
\label{f_eta-}
    f_{\eta}^{(\b<0)}(r) \in span \Big\{& r^{\frac{1}{2} \left(2-\eta\pm\sqrt{9 \eta ^2+\frac{24}{\eta -2}+16}\right)}, 
    r^{\frac{1-\eta }{2}} e^{\pm r\sqrt{\frac{(\eta -1) \eta +1}{-\beta }}} \Big\}, \\
\label{f_den+-}
    f_{den}(r) \in span \Big\{ &e^{-s r^2} : s=\frac{3\eta}{2\gamma}, \frac{3\eta^2}{4\gamma(2-\eta)}, \frac{3\eta}{8\gamma}\left(1\pm\sqrt{1+\frac{8\gamma}{3\beta}}\right) \Big\}.
\end{align}

We discuss the results. 
First, each of the solution spaces is four dimensional, because we are considering fourth-order differential equations.
Second, the first two bases in each solution space are independent of $\b$. 
In fact, they can be obtained from the equation of motion for $\b=0$ \eqref{eom0} in a similar procedure, and \eqref{eom0} is a second-order differential equation.
Third, depending on the sign of $\b$, 
the behavior of the flows around each asymptotic solution changes drastically. 

Let's take a closer look. 

For $f_{S}^{(\b>0)}(r)$, we have three marginal directions $\{1, e^{\pm i \f{r}{\sqrt{\b}}}\}$ and a relevant one $\{r^{\frac{\eta }{2-\eta }}\}$ 
because $\frac{\eta }{2-\eta }>0$ for \eqref{eta}.
For $f_{S}^{(\b<0)}(r)$, on the other hand, we have a marginal one $\{1\}$, two relevant ones $\{r^{\frac{\eta }{2-\eta }}, e^{+\f{r}{\sqrt{-\b}}}\}$, 
and an irrelevant one $\{e^{-\f{r}{\sqrt{-\b}}}\}$.

For $f_{\eta}^{(\b>0)}(r)$, the first two directions are irrelevant, while the latter two are marginal. 
For $f_{\eta}^{(\b<0)}(r)$, only \{$r^{\frac{1-\eta}{2}} e^{+r\sqrt{\frac{(\eta-1)\eta+1}{-\beta}}}$\} is relevant, and all other directions are irrelevant.

For $f_{den}(r)$ with $\b>0$, $e^{-sr^2}$ with $s=\frac{3\eta}{8\gamma}\left(1-\sqrt{1+\frac{8\gamma}{3\beta}}\right)<0$ 
is relevant because of $\g >0$, while the other three are irrelevant. 
For $\b<0$, all four directions are irrelevant, and thus $a_{den}(r)$ is the most attractive. 

Later, this information will be summarized in flowcharts, Figs. \ref{f:flow0}, \ref{f:flow+}, and \ref{f:flow-}.

\subsection{Global Structure of the Solution Space}\label{s:global}
Next, we study the global structure of the solution space by checking the flows numerically. 
Here we consider only solutions with non-negative $a(r)$.
Before beginning a detailed analysis, we consider the effect of $\b \Box R$ in advance. 


As shown in Sec.~\ref{s:sol}, the self-consistent equation for $a(r)$ \eqref{eom} with $\b\neq0$ contains three asymptotic solutions\footnote{Note that we are neglecting the negative energy metric $a_{neg}(r)$.},
while for $\b=0$ two belong to the $\ME_-$ branch and the other belongs to the $\ME_+$ branch. 
This should mean that it is the higher-order differential term $\b \Box R$ that connects 
the two branches in the solution space. 
In fact, it is well known that transitions between solutions of different branches can be caused by a higher-order differential term \cite{singularPerturbation}. 
In our case, this means that $\beta$ provides flows between $\ME_+$ and $\ME_-$ branches.
Furthermore, our results in Sec.~\ref{s:local} indicate that the structure of the global flow between $\ME_+$ and $\ME_-$ branches depends on the sign of $\b$. 
In the following, we will see that these predictions are indeed the case.

\subsubsection*{The flow for $\b=0$}
We start with the case of $\beta=0$, where we have two branches of equations: $\ME_+=0$ and $\ME_-=0$, \eqref{eom0}. 

\begin{figure}[!ht] 
    \centering
    \includegraphics[scale =0.6]{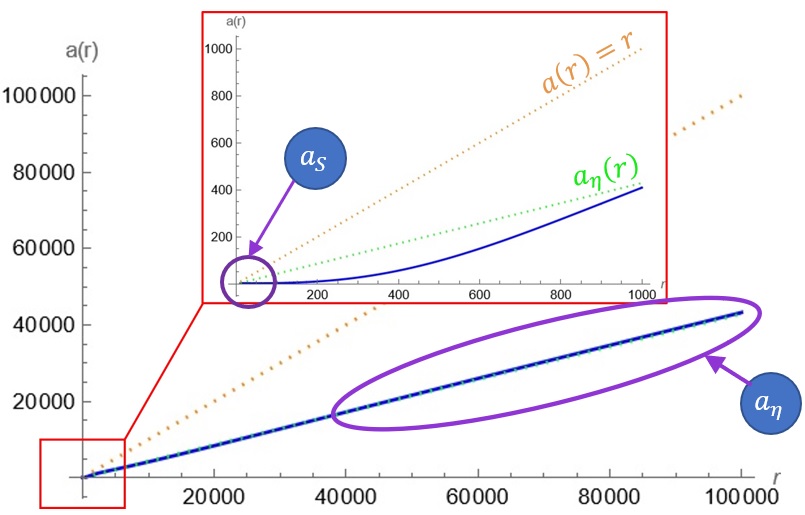}
    \caption{Flow (a). 
     Typical behavior of $a(r)$ going from $a_{S}(r)$ to $a_\eta(r)$.}
    \label{f:a}
\end{figure}

We first examine $\ME_-=0$, which includes $a_S(r)$ and $a_\eta(r)$.
From the numerical calculations given by Fig.~\ref{f:a}, 
we can see that it flows from $a_S(r)$ to $a_{\eta}(r)$, and we call it flow (a). 
This is true for any initial values obtained by perturbing $a_S(r)$. 
Therefore, the basis $\{1\}$ in $f_S(r)$ \eqref{f_S+} is marginally relevant. 
Note that this flow is insensitive to the value of $\beta$.

\begin{figure}[!htb]
    \begin{minipage}[!ht]{0.45\linewidth}
        \centering
        \includegraphics[keepaspectratio, scale=0.6]{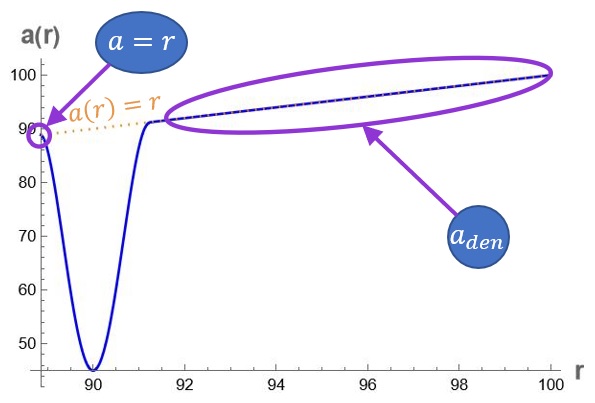}
        \caption{Flow (b). 
        Typical behavior of $a(r)$ that connects two irrelevant directions of $a_{den}(r)$ to $a=r$ singularity.
        }
        \label{f:b}
    \end{minipage}
    \hfill
    \begin{minipage}[!ht]{0.45\linewidth}
        \centering
        \includegraphics[keepaspectratio, scale=0.5]{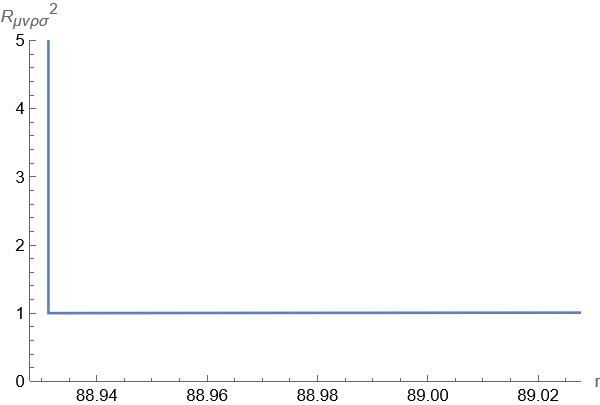}
        \caption{$R^{\mu\nu\a\b}R_{\mu\nu\a\b}$ for Fig.~\ref{f:b} 
        diverges at the point satisfying $a(r)=r$.}
        \label{f:singular}
    \end{minipage}
\end{figure}

We next consider $\ME_+=0$.
This contains only $a_{den}(r)$, \eqref{a_dena}, 
which has two irrelevant directions (the first two in \eqref{f_den+-}).
Therefore, starting from $a_{den}(r)$ and solving $\ME_+=0$ in the direction of decreasing $r$, 
we should always move away from $a_{den}(r)$, regardless of the two initial values.
In fact, numerical calculations show that this is the case (see Fig.~\ref{f:b}.). 
Moreover, they also show that 
after moving away, it approaches a physical singularity at a point satisfying $a(r)=r$ (See Fig.~\ref{f:singular}.). 
Indeed, the geometric invariants such as $R^{\mu\nu\a\b}R_{\mu\nu\a\b}$ contain the factor $r-a(r)$ in the denominator. 
We call the flow from $a=r$ to $a=a_{den}(r)$ flow (b).
We also find it insensitive to the value of $\beta$.
Note that the part with a negative slope in Fig.~\ref{f:b} has a negative energy density
(since $\bra -T^t{}_t\ket = \f{1}{8\pi Gr^2}\p_r a(r)$ from \eqref{Mr}), 
which may give rise to a trapped surface there \cite{Ho_nega}. 
This may have something to do with the occurrence of singularities. 

\begin{figure}[!htb] 
    \centering
    \includegraphics[scale =0.6]{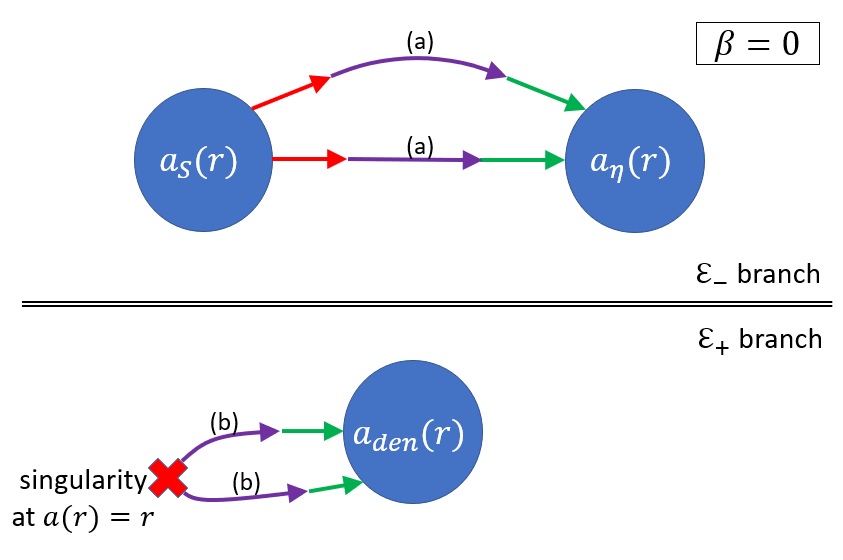}
    \caption{The global structure of the solution space for $\b=0$.
    Red (green) arrows represent relevant (irrelevant) directions.
    The global flows (a) and (b) correspond to Fig.~\ref{f:a} and Fig.~\ref{f:b}, respectively.}
    \label{f:flow0}
\end{figure}

Thus, we obtain the structure of the global flow in the solution space for $\b=0$, where the two branches are in general disconnected. See Fig.~\ref{f:flow0}. 

\newpage
\subsubsection*{The flow for $\b>0$}

\begin{figure}[!ht]
  \begin{minipage}[!ht]{0.55\linewidth}
    \centering
    \includegraphics[keepaspectratio, scale=0.35]{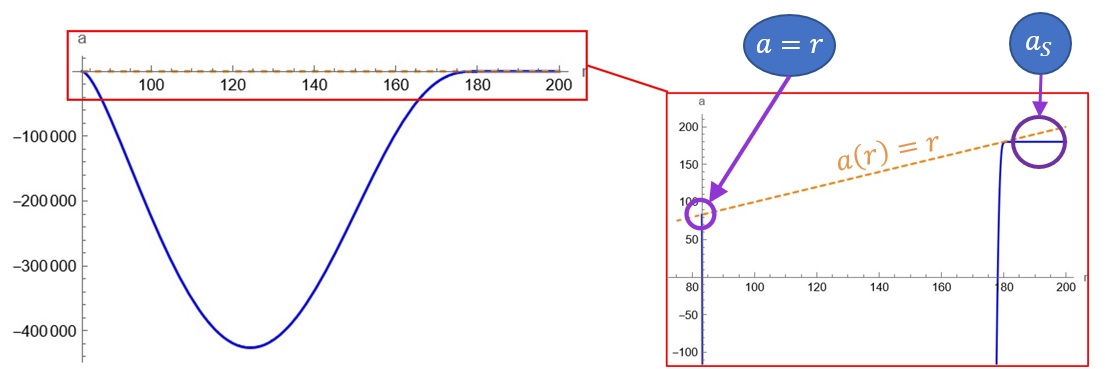}
    \caption{Flow (c).}
    \label{f:c}
  \end{minipage}
  \hfill
  \begin{minipage}[!ht]{0.45\linewidth}
    \centering
    \includegraphics[keepaspectratio, scale=0.4]{imgd.jpg}
    \caption{Flow (d).}
    \label{f:d}
  \end{minipage}
\end{figure}

\begin{figure}[!ht]
  \begin{minipage}[!ht]{0.55\linewidth}
    \centering
    \includegraphics[keepaspectratio, scale=0.35]{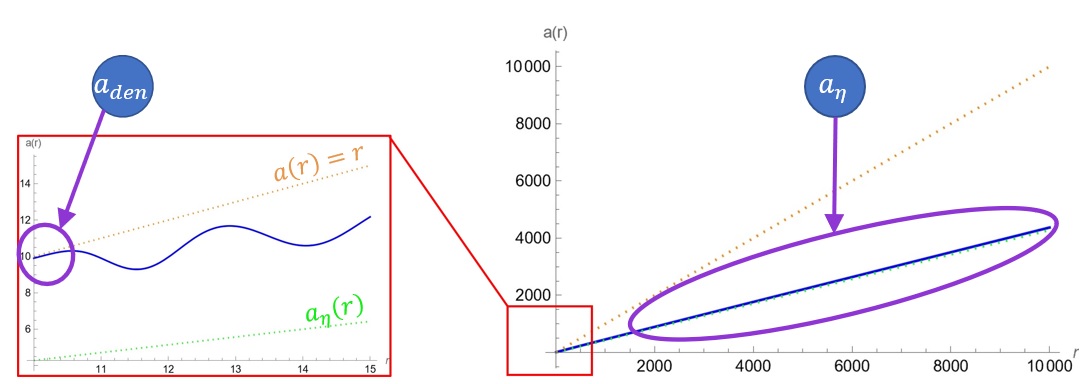}
    \caption{Flow (e).}
    \label{f:e}
  \end{minipage}
  \hfill
  \begin{minipage}[!ht]{0.45\linewidth}
    \centering
    \includegraphics[keepaspectratio, scale=0.37]{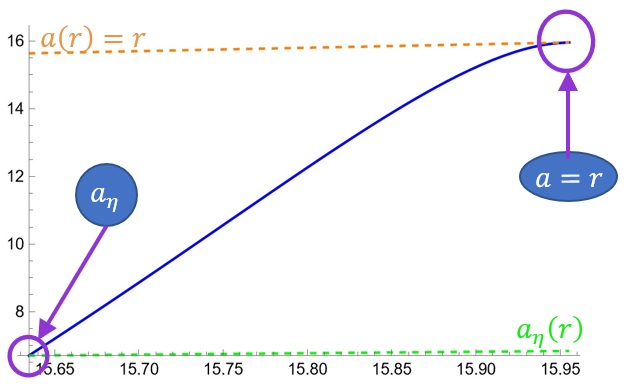}
    \caption{Flow (f).}
    \label{f:f}
  \end{minipage}
\end{figure}

\begin{figure}[!ht] 
    \centering
    \includegraphics[scale =0.6]{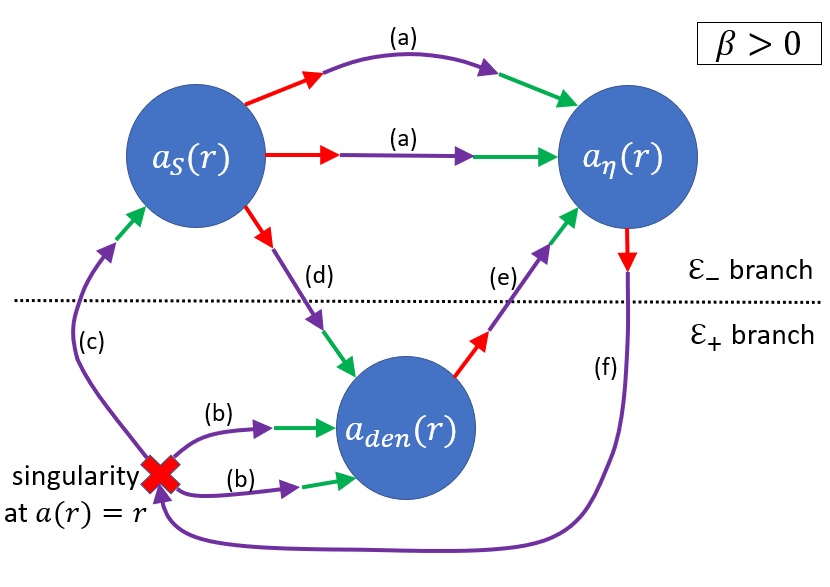}
    \caption{The global structure of the solution space for $\b>0$.}
    \label{f:flow+}
\end{figure}

We now investigate \eqref{eom} when $\b>0$ numerically. 
In this case, we have 4 new kinds of flows due to $\b\neq 0$, which are shown in Figs. \ref{f:c} -- \ref{f:f} and summarized in Fig.~\ref{f:flow+}.


Let's start with two flows related to $a_S(r)$ from its marginal directions $e^{\pm i \frac{r}{\sqrt{\beta}}}$ in \eqref{f_S+}.
It turns out that one of the directions is (marginally) irrelevant (see Fig.~\ref{f:c}, which we call flow(c).) and the other direction is (marginally) relevant (see Fig.~\ref{f:d}, which we call flow(d).).
Fig.~\ref{f:c} represents a flow from $a=r$ singularity to $a_S(r)$.
Fig.~\ref{f:d} represents a flow from $a_S(r)$ to $a_{den}(r)$ through its irrelevant direction $\frac{3\eta}{8\gamma}\left( 1+\sqrt{1+\frac{8\gamma}{3\beta}} \right)$ in \eqref{f_den+-}.

The other two new flows are related to $a_\eta(r)$. 
One flows from $a_{den}(r)$ to $a_\eta(r)$ (see Fig.~\ref{f:e}\footnote{After deviating from $a_{den}(r)$, it oscillates due to the higher derivative terms as in Fig.~\ref{f:e}. Similar behaviors appear in different approaches to quantum black holes \cite{Zhang:2014bea,Bonanno:2019rsq,Casadio:2021eio,Borissova:2022mgd}.}, which we call flow (e).),
while the other one flows from $a_\eta(r)$ to a singularity, $a(r) = r$, (see Fig.~\ref{f:f}, which we call flow (f).).
Note that although Fig.~\ref{f:f} seems to show that it is approaching $a_{den}(r)$, it is not\footnote{In the case of flowing into $a_{den}(r)$, we can always increase the precision so that it is hitting singularity at larger $r$, but it cannot be done in the case here.};
it always hits the singularity within a short interval of $r$.

The resulting flowchart for the case $\b>0$ is shown in Fig.~\ref{f:flow+}.

\subsubsection*{The flow for $\b<0$}

\begin{figure}[!ht]
  \begin{minipage}[!ht]{0.45\linewidth}
    \centering
    \includegraphics[keepaspectratio, scale=0.5]{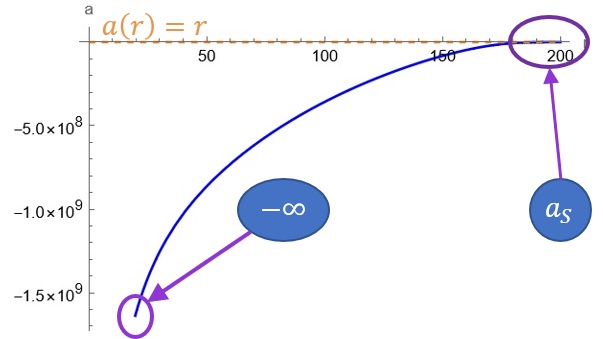}
    \caption{Flow (g).}
    \label{f:g}
  \end{minipage}
  \hfill
  \begin{minipage}[!ht]{0.45\linewidth}
    \centering
    \includegraphics[keepaspectratio, scale=0.5]{imgh.jpg}
    \caption{Flow (h).}
    \label{f:h}
  \end{minipage}
\end{figure}
\begin{figure}[!ht]
  \begin{minipage}[!ht]{0.45\linewidth}
    \centering
    \includegraphics[keepaspectratio, scale=0.5]{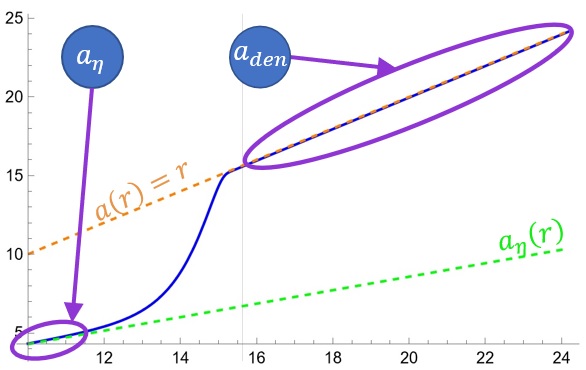}
    \caption{Flow (i).}
    \label{f:i}
  \end{minipage}
  \hfill
  \begin{minipage}[!ht]{0.45\linewidth}
    \centering
    \includegraphics[keepaspectratio, scale=0.5]{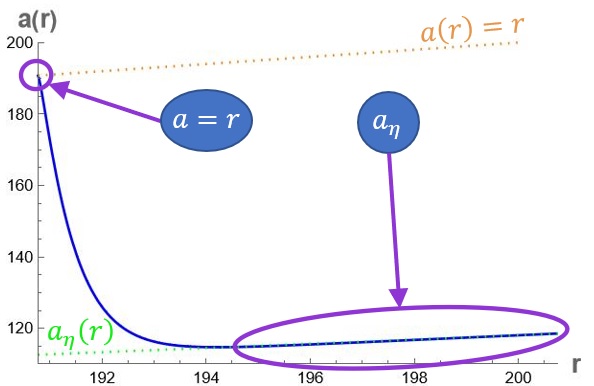}
    \caption{Flow (j).}
    \label{f:j}
  \end{minipage}
\end{figure}
\begin{figure}[!ht] 
    \centering
    \includegraphics[scale =0.55]{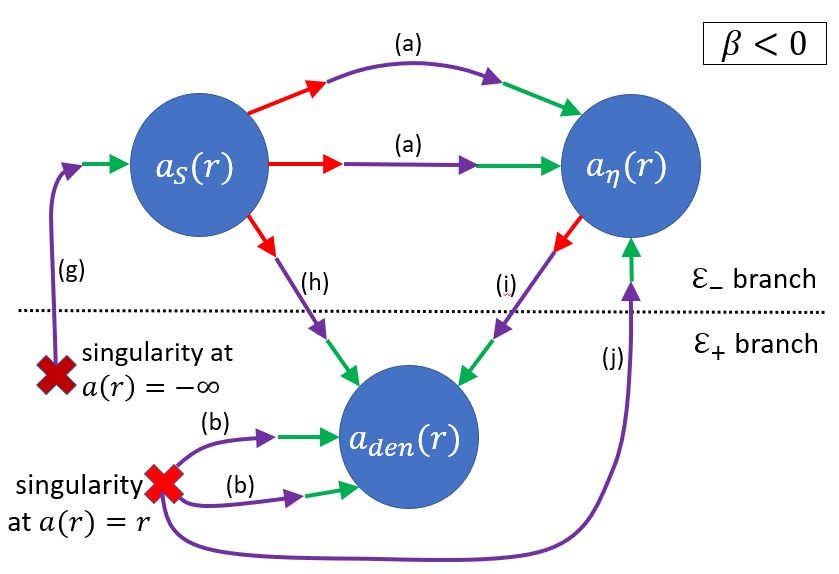}
    \caption{The global structure of the solution space for $\b<0$.}
    \label{f:flow-}
\end{figure}

We finally examine the flow for $\b<0$. 
In this case, the flows (c), (d), (e), (f) in Fig.~\ref{f:flow+} are replaced by the flows (g), (h), (i), (j) in Fig.~\ref{f:flow-}. 
They are shown in Figs.\ref{f:g} -- \ref{f:j}.

Two of them are related to $a_S(r)$. 
Flow (g) in Fig.~\ref{f:g} represents a flow from negative infinity to $a_S(r)$ through its irrelevant direction, $e^{-\frac{r}{\sqrt{-\b}}}$ in \eqref{f_S-}.
Flow (h) in Fig.~\ref{f:h} represents a flow from $a_S(r)$ through its relevant direction, $e^{\frac{r}{\sqrt{-\b}}}$ in \eqref{f_S-}, to $a_{den}(r)$.


The other two flows are related to $a_\eta(r)$:
they are in opposite directions of the $\b>0$ case. 
For example, when $\b>0$, the direction of the flow is from $a_{den}(r)$ to $a_{\eta}(r)$ as in Fig.~\ref{f:e}, but now it is from $a_\eta(r)$ to $a_{den}(r)$ (see Fig.~\ref{f:i}, which we call flow (i).). 
A similar thing happens to the flow between the $a_\eta(r)$ and the $a=r$ singularity. 
Instead of a flow to $a=r$ singularity, we have a flow from $a=r$ singularity to $a_\eta(r)$ (see Fig.~\ref{f:j}, which we call flow (j)). 

We then summarize the results in Fig.~\ref{f:flow-}. 

\newpage


\section{Solutions for Linearized equation of motion}\label{A:f}
In this appendix, we demonstrate how one can derive the solutions of linearized equation, i.e., how to obtain \eqref{f_S+} - \eqref{f_den+-} from \eqref{f_S} - \eqref{f_den}.

Let us start from the linearized equations \eqref{f_S} -- \eqref{f_den}, which are 
\begin{align}
    \label{f_S_a}
    \b f_{S}''''(r)
    &-\frac{2 \b(\eta -1) }{(2-\eta ) r} f_{S}'''(r)
    +f_{S}''(r)
    -\frac{2 (\eta-1)}{(2-\eta ) r}f_{S}'(r) = 0 
    \\
    \label{f_eta_a}
    \b f_{\eta}''''(r)
    &+\frac{2 \b (\eta -1) }{r}f_{\eta}'''(r)
    +\left(\eta ^2-\eta+1\right)f_{\eta}''(r) \nonumber\\
    &+\frac{(\eta -1) \left(\eta ^2-\eta+1\right)}{r}f_{\eta}'(r)
    +\frac{\eta (\eta -1)(2 \eta -1)\left(\eta ^2-\eta +1\right) }{(2-\eta ) r^2}f_{\eta}(r) = 0
    \\
    \label{f_den_a}
    \b \bar f_{den}''''\left(r^2\right)
    &+\frac{3 \b (3-\eta ) \eta}{2 \gamma  (2-\eta)}\bar f_{den}'''\left(r^2\right)
    -\frac{3 \eta ^2 (2\gamma  (2-\eta )-3 \beta  (\eta +4))}{16\gamma^2 (2-\eta)}\bar f_{den}''\left(r^2\right)\nonumber\\
    &-\frac{9 \eta ^3 (\gamma  (4-\eta )-3\beta \eta)}{32 \gamma ^3 (2-\eta)}\bar f_{den}'\left(r^2\right)
    -\frac{27 \eta ^5}{64 \gamma ^3(2-\eta)}\bar f_{den}\left(r^2\right) = 0.
\end{align}

\subsection*{$f_S$-equation}

First of all, the $f_S$-equation \eqref{f_S_a} can be rewritten as 
\begin{align}
    0
    &=\left(\left(D -\frac{2 (\eta -1) }{(2-\eta ) r}\right)\b D^3 + \left(D -\frac{2 (\eta -1) }{(2-\eta ) r}\right)D \right)f_S(r) \\
    &=\left(D -\frac{2 (\eta -1) }{(2-\eta ) r}\right) \left(\b D^2 + 1 \right) D f_S(r),
\end{align}
where $D=\frac{\text{d}}{\text{d} r}$. The equation then can be solved by hand in this form from the left-most operator, $\left(D -\frac{2 (\eta -1) }{(2-\eta ) r}\right)$, to the right. Then we obtain
\begin{align*}
    f_{S}^{(\b>0)}(r) \in span \Big\{ &1, r^{\frac{\eta }{2-\eta }}, e^{\pm i \f{r}{\sqrt{\b}}} \Big\}, \\
    f_{S}^{(\b<0)}(r) \in span \Big\{ &1, r^{\frac{\eta }{2-\eta }}, e^{\pm \f{r}{\sqrt{-\b}}} \Big\}.
\end{align*}
Here we separate them into $\b>0$ and $\b<0$ cases for real behavior over the square root.

\subsection*{$f_\eta$-equation}

What we consider next is the $f_\eta$-equation \eqref{f_eta_a}. 
In the following, we are going to solve it in two steps. 
First, we solve the $f_\eta$-equation partially to find a suitable ansatz for the form of solution. 
Then, when we fix the ansatz, we can solve the $f_\eta$-equation perturbatively.

As our first step, at $r \rightarrow \infty$, we select out terms from \eqref{f_eta_a} by the dimensions of coefficients in $r$, and it gives 
\begin{align}
    \b f_{\eta}''''(r)
    +\left(\eta ^2-\eta+1\right)f_{\eta}''(r) \nonumber= 0,
\end{align}
which can be solved by transforming it into a characteristic equation. 
With $f_\eta(r) = e^{\lambda r}$, we get $\lambda = 0,0,\pm\sqrt{\frac{\eta ^2-\eta+1}{-\b}}$. 
This incentivizes us to solve the equation perturbatively with powers of $r$ and $e^{\pm\sqrt{\frac{\eta ^2-\eta+1}{-\b}}r}$. 
Therefore, we set the ansatz
\begin{align}
    f_\eta(r) 
    = r^k e^{\pm\sqrt{\frac{\eta ^2-\eta+1}{-\b}} r}\left( 1+\frac{c_1}{r}+\dots \right).
\end{align}
By inserting this ansatz into \eqref{f_eta_a}, we can obtain the full form of $f_\eta$ perturbatively with respect to the order of $r$. As an example, we can solve the first two leading orders to obtain 
\begin{align}
    k &= \frac{1-\eta}{2}, \\ 
    c_1 &= \frac{(\eta -1) (4 \beta  (\eta -6) (\eta -2)+((\eta -1) \eta +1) (\eta  (7 \eta -11)+18))}{8 \beta  (\eta -2) \sqrt{\frac{-\eta ^2+\eta
   -1}{\beta }}}.
\end{align}

As for the power-like solutions, we can use the ansatz
\begin{align}\label{f_eta_ansatz_2}
    f_\eta(r) = r^k \left( 1 + \frac{d_2}{r^2} + \dots  \right).
\end{align}
Note that the corrections are set to respect the $Z_2$ symmetry, $r \leftrightarrow (-r)$, of \eqref{f_eta_a}. We then again solve it perturbatively after inserting it into the \eqref{f_eta_a}. Note that since this ansatz is for both zero $\lambda$'s, we expect to have two values for $k$ in \eqref{f_eta_ansatz_2}; indeed, that is the case, and we label them $k_+$ and $k_-$. These two cases up to the first two leading orders are
\begin{align}
    k_- &= \frac{1}{2} \left(2-\eta - \sqrt{9 \eta ^2+\frac{24}{\eta -2}+16}\right), \\
    d_{2,-} &= \beta  (\eta -1) \eta \frac{d_{2,-,n}}{d_{2,-,d}}, \\
    d_{2,-,n} &= -3 \eta ^4-19 \eta ^3+60 \eta ^2-48 \eta +20-(2-\eta ) (\eta^2 -13\eta +10) \sqrt{9 \eta ^2+\frac{24}{\eta -2}+16},  \\
    d_{2,-,d} &= 4 (2-\eta)^2 (\eta^2 -\eta +1) \left(2+\sqrt{9 \eta ^2+\frac{24}{\eta -2}+16}\right),
\end{align}
and 
\begin{align}
    k_+ &= \frac{1}{2} \left(2-\eta + \sqrt{9 \eta ^2+\frac{24}{\eta -2}+16}\right), \\
    d_{2,+} &= \beta  (\eta -1) \eta \frac{d_{2,+,n}}{d_{2,+,d}}, \\
    d_{2,+,n} &= 3 \eta ^4+19 \eta ^3-60 \eta ^2+48 \eta -20-(2-\eta) (\eta ^2-13 \eta +10) \sqrt{9 \eta ^2+\frac{24}{\eta -2}+16},  \\
    d_{2,+,d} &= 4 (2-\eta )^2 (\eta^2 -\eta +1) \left(-2+\sqrt{9 \eta ^2+\frac{24}{\eta -2}+16}\right).
\end{align}
As a result, we have $f^{(\b<0)}_\eta(r)$ to be spanned by the following 4 bases at leading order 
\begin{align}
    f_{\eta}^{(\b>0)}(r) \in span \Big\{& r^{\frac{1}{2} \left(2-\eta\pm\sqrt{9 \eta ^2+\frac{24}{\eta -2}+16}\right)}, 
    r^{\frac{1-\eta }{2}} e^{\pm i r\sqrt{\frac{(\eta -1) \eta +1}{\beta }}} \Big\}, \\
    f_{\eta}^{(\b<0)}(r) \in span \Big\{& r^{\frac{1}{2} \left(2-\eta\pm\sqrt{9 \eta ^2+\frac{24}{\eta -2}+16}\right)}, 
    r^{\frac{1-\eta }{2}} e^{\pm r\sqrt{\frac{(\eta -1) \eta +1}{-\beta }}} \Big\}.    
\end{align}

\subsection*{$f_{den}$-equation}

The $f_{den}$ case is the easiest one to solve since it can be directly solved with the ansatz for characteristic equation, $f_{den}(t)=e^{-st}$, of \eqref{f_den_a}. The solutions are 
\begin{align}
    s = \frac{3\eta}{2\gamma}, \frac{3\eta^2}{4\gamma(2-\eta)}, \frac{3\eta}{8\gamma}\left(1\pm\sqrt{1+\frac{8\gamma}{3\beta}}\right).
\end{align}
Hence, 
\begin{align}
    f_{den}(r) \in span \Big\{ &e^{-s r^2} : s=\frac{3\eta}{2\gamma}, \frac{3\eta}{4\gamma(2-\eta)}, \frac{3\eta}{8\gamma}\left(1\pm\sqrt{1+\frac{8\gamma}{3\beta}}\right) \Big\}.
\end{align}


\end{document}